\documentclass[11pt]{article}
\usepackage[utf8]{inputenc}

\usepackage{amsmath,fullpage,graphicx,color,amsfonts,url,multirow,amssymb,tabularx,setspace}
\usepackage{natbib}
\usepackage{multibib}
\usepackage{chngcntr}
\usepackage{sectsty}
\usepackage[top=.75in,bottom=.75in,left=.75in,right=.75in]{geometry}
\usepackage[pdftex,breaklinks]{hyperref}

\sectionfont{\fontsize{12}{15}\selectfont}
\subsectionfont{\fontsize{12}{15}\selectfont}

\newcites{SM}{Supplemental references}

\newcommand\independent{\protect\mathpalette{\protect\independenT}{\perp}}
\def\independenT#1#2{\mathrel{\rlap{$#1#2$}\mkern2mu{#1#2}}}

\newcolumntype{L}[1]{>{\raggedright\arraybackslash}p{#1}}
\newcolumntype{C}[1]{>{\centering\arraybackslash}p{#1}}
\newcolumntype{R}[1]{>{\raggedleft\arraybackslash}p{#1}}

\title{\LARGE Integrated causal-predictive machine learning models for tropical cyclone epidemiology}
\author{\normalsize Rachel C. Nethery$^1$, Nina Katz-Christy$^2$, Marianthi-Anna Kioumourtzoglou$^3$,\\
\normalsize Robbie M. Parks$^3$, Andrea Schumacher$^4$, G. Brooke Anderson$^5$\\

\\

\small $^1$Department of Biostatistics, Harvard T.H. Chan School of Public Health\\
\small $^2$Department of Statistics, Harvard University, Faculty of Arts and Sciences\\
\small $^3$Department of Environmental Health Sciences, Columbia Mailman School of Public Health\\
\small $^4$Cooperative Institute for Research in the Atmosphere, Colorado State University\\
\small $^5$Department of Environmental \& Radiological Health Sciences, Colorado State University}
\date{}

\begin{document}

\maketitle

\begin{abstract}
    Strategic preparedness has been shown to reduce the adverse health impacts of hurricanes and tropical storms, referred to collectively as tropical cyclones (TCs), but its protective impact could be enhanced by a more comprehensive and rigorous characterization of TC epidemiology. To generate the insights and tools necessary for high-precision TC preparedness, we develop and apply a novel Bayesian machine learning approach that standardizes estimation of historic TC health impacts, discovers common patterns and sources of heterogeneity in those health impacts, and enables identification of communities at highest health risk for future TCs. The model integrates (1) a causal inference component to quantify the immediate health impacts of recent historic TCs at high spatial resolution and (2) a predictive component that captures how TC meteorological features and socioeconomic/demographic characteristics of impacted communities are associated with health impacts. We apply it to a rich data platform containing detailed historic TC exposure information and Medicare claims data. The health outcomes used in our analyses are all-cause mortality and cardiovascular- and respiratory-related hospitalizations. We report a high degree of heterogeneity in the acute health impacts of historic TCs at both the TC level and the community level, with substantial increases in respiratory hospitalizations, on average, during a two-week period surrounding TCs. TC sustained windspeeds are found to be the primary driver of increased mortality and respiratory risk. Our modeling approach has broader utility for predicting the health impacts of many types of extreme climate events.
\end{abstract}

\section{Introduction}\label{s:intro}
%Tropical cyclones (TC), also known as hurricanes, are rotating low-pressure weather systems that develop in seven tropical ocean basins worldwide and regularly impact at least 33 nations. TC are characterized by the severe winds, rainfall, and flooding they often bring \citep{shultz2005epidemiology}. Atlantic basin TC regularly hit the eastern and southern coastlines of the US; in recent years, Hurricanes Harvey, Florence, Michael, and Maria have devastated communities in these regions. 
%Wind and flooding from these TC have caused massive property destruction and infrastructure damage, substantial population displacement, and severe economic hardship in many communities.
The US National Oceanic and Atmospheric Administration reports that tropical cyclones (TCs) impose the largest financial burden of any weather disasters in the US, costing \$945.9 billion since 1980 or roughly \$21.5 billion per event \citep{noaa2020}. TCs, which include hurricanes and tropical storms, often bring severe winds, rainfall, and flooding \citep{shultz2005epidemiology}, which can catalyze massive property and infrastructure damage. Due to the diverse types of hardships that can be set in motion by TC, the full spectrum of human health impacts of TCs are incompletely understood and unreliably quantified. Extreme weather events are known to cause both ``direct'' and ``indirect'' health impacts. It is well-appreciated that TCs introduce severe risks for accidental mortality and injuries \citep{centers2005surveillance, rappaport2014fatalities,rappaport2016fatalities,gong2007injuries}, such as drowning or blunt force trauma from falling debris, which are known as ``direct'' health impacts, as they are straightforwardly attributed to the TC (i.e., the causal mechanism has been clearly identified). Direct TC health impacts are generally the focus of post-storm surveillance in the US.

TCs can also ``indirectly'' elevate risk for a range of other adverse health events because, for example, they often cause power outages \citep{chen2018impacts, han2009, klinger2014power,lane2013health, sanchez2017cnn, shultz2017preparing}, trigger mass evacuations \citep{lew1996mortality, brunkard2008hurricane,centers2013deaths, dosa2012evacuate}, create psychological stress \citep{lutgendorf1995physical, lenane20172115}, require clean-up \citep{centers2004preliminary,lew1996mortality}, increase exposure to heat and pollution \citep{shultz2017preparing}, and interfere with normal medical care and medication use \citep{klinger2014power,gray2007hospitals}.
%not sure if we have citations or space for these but TCs also: disrupt access to food, (potentially) increase crime, affect substance abuse, impact spreads of infectious diseases.
Post-storm surveillance can hugely underestimate these indirect health impacts of TCs, as evidenced by Hurricane Maria. While surveillance initially attributed 64 deaths in Puerto Rico to the storm \citep{puerto2018}, later epidemiological studies estimated that the storm caused $>$2,000 deaths \citep{kishore2018maria, santos2018acertainment,rivera2018estimating, santos2018use,puerto2018}.
%In these studies, mortality during/after the storm is compared with expected mortality based on trends in prior years to estimate the excess deaths caused by the storm. Several studies of this kind have been conducted to better estimate indirect TC health impacts but have largely been limited to a single storm \citep{sharma2008chronic,mcquade2018emergency, gotanda2015hurricane, swerdel2014effect,greenstein2016impact, kim2016effect} or year \citep{mckinney2011direct}.

%Because warm sea surface temperatures are often associated with more intense TC, \citep{shultz2005epidemiology,sobel2016human,xu2016relationship} climate change is projected to increase average TC intensity over this century \citep{knutson2015global,knutson2019tropical,knutson2020tropical}, although some uncertainty remains. Meanwhile, population growth and development in coastal areas is bringing more people into the direct path of TCs. Thus, TCs are expected to present an escalating threat to human health in the US in the coming years. Strategic storm preparedness in response to an impending TC is believed to be one of the most effective tactics for minimizing TC health impacts \citep{thomalla2004we,shultz2005epidemiology,keim2008building}. Minimizing the threat of TC to human health in the US will require new, high-precision tools to inform strategic storm preparedness.

The literature on TC epidemiology has been dominated by single-storm studies \citep{sharma2008chronic,mcquade2018emergency, gotanda2015hurricane, swerdel2014effect,greenstein2016impact, kim2016effect}, seeking to quantify the total excess mortality or morbidity caused by a TC (including both direct and indirect effects). This focus on single storms was driven by widespread acknowledgement of substantial heterogeneity in TC health impacts. Recently, the first large-scale study was conducted to estimate average health effects across all TCs impacting the US over a 12-year period \citep{yan2020tropical}. While these studies have helped to reveal fundamental features of TC epidemiology, the results of single-storm studies may not generalize well, and multi-storm average health effects are too coarse to explain across-storm variability. Thus these studies have been unable deliver the targeted yet generalizable insights needed to guide strategic storm preparedness, which is believed to be one of the most effective tactics for minimizing TC health impacts \citep{thomalla2004we,shultz2005epidemiology,keim2008building}. A 2020 report by the National Academies of Sciences, Engineering, and Medicine stressed that in order to strengthen disaster resilience, improve responses, and quicken recoveries, the US needs a uniform approach to quantifying disaster-related mortality and morbidity, as well as new analytical methods to enable estimation of disaster health impacts and the capacity to implement such methods on population-level data \citep{nas2020framework}.

%The goal of our work is to create a predictive tool that provides information in real time about the areas of highest health risk and the types of acute health risks anticipated for an impending TC, in order to maximize the protective impact of strategic preparedness efforts. In the US, hurricane forecasts are generally known with some degree of certainty several days in advance of the storm's first landfall. These forecasts include information about the storm's expected path and the expected wind speeds, rainfall amounts, and flooding for communities on that path. Throughout this paper, we make the assumptions that these forecasted features of the approaching TC are fixed and known. Given these characteristics, we wish to predict the health risks it is expected to cause in the communities on its path.

The goal of our work is to inform strategic TC preparedness through development and application of a new modeling approach that (1) standardizes estimation of acute health impacts across past TCs, (2) discovers common patterns and sources of heterogeneity in those health impacts, and (3) enables identification of communities at highest health risk for future TCs.
%Constructing a model for real-time, spatially-granular prediction of the acute health impacts of an impending TC presents two main challenges.
%Due to the absence of reliable data about the total (direct and indirect) health effects of past TCs, the first obstacle to constructing our predictive model is, in fact, a causal inference problem.
First, the proposed approach must incorporate a causal inference component that, when applied to historic data, estimates the excess adverse health events caused by past TCs (hereafter ``health effects'' or ``health impacts'') at high spatial resolution in a standardized and transparent fashion. These estimates should capture both direct and indirect effects and should be adjusted for confounding. A TC's health impact in a particular community may be influenced (i.e., modified) by a complex interplay among the features of the storm and the population  \citep{keim2008building}, and understanding these drivers of heterogeneity is a key aim of our work. Thus, the second component of our approach is a predictive model relating the community- and TC-specific health impacts to the TC's meteorological features and the socioeconomic/demographic features of the community. In addition to offering unprecedented insights into multi-storm TC epidemiology, this approach allows for community-specific prediction of the health impacts of an approaching TC with forecasted track and features. The predictive model could also be used to create general community-level TC health risk profiles based on a collection of representative future TC exposures. This tool represents a first step toward identifying communities at highest risk for adverse TC health impacts so that they can be targeted for immediate TC strategic preparedness and/or long-term efforts to increase resilience.

Building on a rich dataset of recent historic US TC exposures and Medicare claims, we introduce an innovative statistical modeling approach that incorporates both the causal inference and predictive components described above. %We seek to characterize how health risks differ during TC exposures compared to non-TC periods (causal inference), as well as how these risks are modified by the characteristics of the affected county and the severity of elements of storm exposure (e.g., wind intensity) within the county (for prediction). 
%We have compiled a unique dataset containing the following information for each county in the eastern US: (1) detailed meteorological features of any TC(s) that affected the county 1999-2015; (2) daily counts of mortality and cause-specific hospitalizations in the Medicare population in the county 1999-2015; and (3) socioeconomic and demographic features of the county.
Our Bayesian machine learning method jointly fits causal inference sub-models to estimate the county-specific health effects of each historic TC, then passes these effect estimates into a predictive sub-model that captures relationships between county and TC features and health impacts. Leveraging recent advances in causal inference with observational pre/post treatment data, the causal sub-models employ a matrix completion approach that adjusts for unmeasured, time-varying confounding under mild assumptions \citep{athey2018matrix}. By joining the causal and predictive models in a Bayesian framework, we account for the uncertainty from all components, and predictions made using this model are accompanied by accurate uncertainty estimates, which are critical to assess their utility. This method can be widely used for characterizing and predicting the health impacts of extreme weather and climate events.

%In Section~\ref{s:methods}, we introduce our historic TC and health records data, formalize the joint causal and predictive Bayesian machine learning model, and describe its implementation on the data. The results of the causal and predictive models are presented in Section~\ref{s:results}, including an evaluation of the predictive performance of the model in held-out samples. In Section~\ref{s:discuss}, we conclude with a discussion of our findings and an explanation of the broader utility of our modeling approach for prediction of the health impacts of future extreme climate events.

\section{Methods}\label{s:methods}

\subsection{Data}\label{s:data}
The data and the study design are described only briefly here, detailed descriptions and justifications of these choices are provided in Section~\ref{s:s:data}. All mortalities, respiratory disease hospitalizations, chronic obstructive pulmonary disease (COPD) hospitalizations, and cardiovascular disease (CVD) hospitalizations in the Medicare population are obtained for the period 1999--2015. Using detailed track and meteorologic data for all Atlantic-basin TCs during the same period, we classify counties as exposed (equivalently ``treated'' for consistency with the causal inference literature) or controls for each TC. Counties that experience TC maximum sustained windspeeds of gale force or higher ($\geq 17.4$ meters/second) at the population mean center are considered treated, and untreated counties within 150 miles of a treated county are eligible to serve as controls. Additional inclusion criteria to prevent modeling instability are applied to determine the final set of analytic treated and control counties for each TC (Section~\ref{s:s:paneldata}).

\begin{figure*}
\centering
\includegraphics[scale=.5]{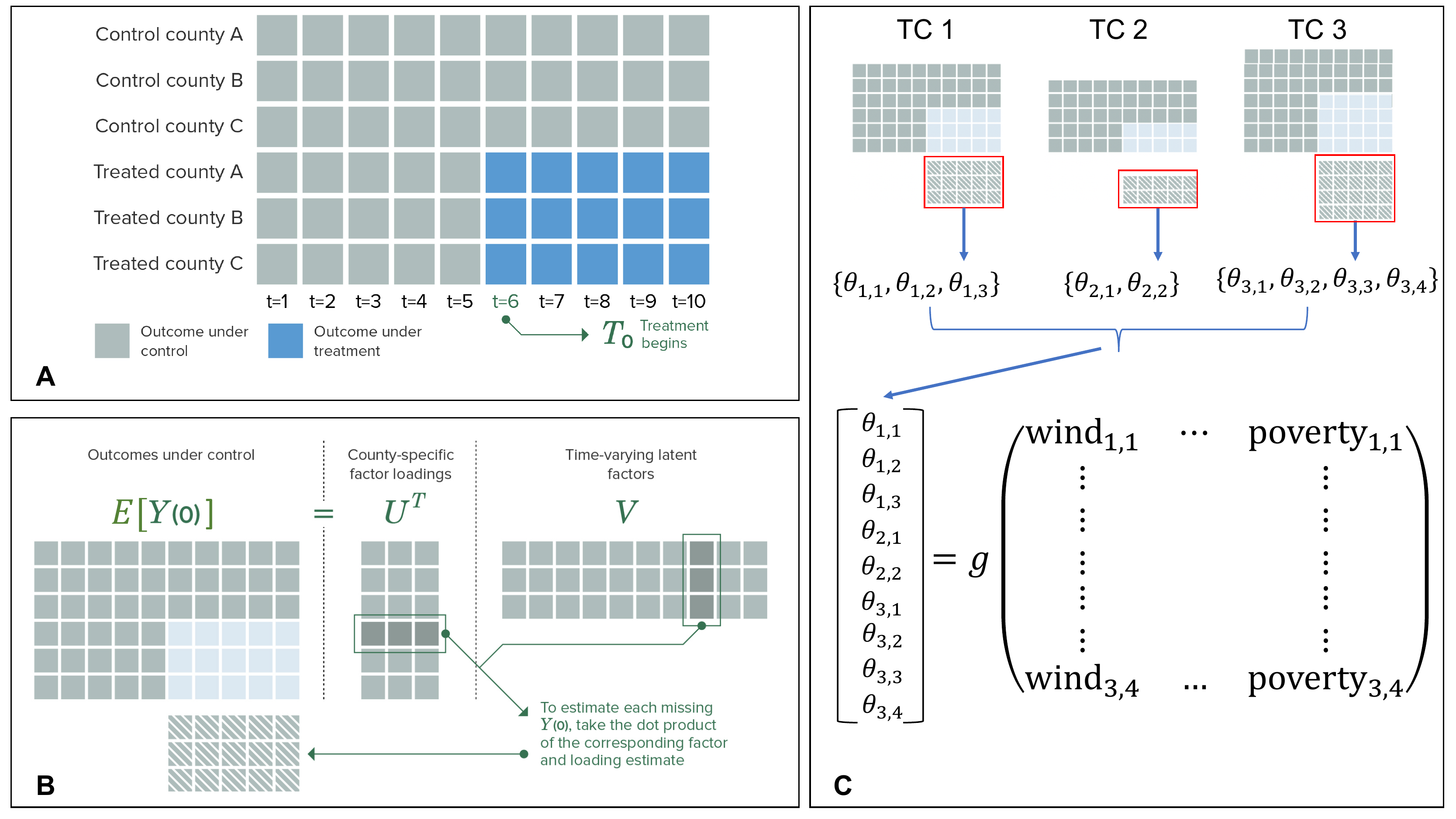}
\caption{Example of panel data structure (A), illustration of matrix completion (B), and visual explanation of our integrated causal and predictive modeling approach (C).}
\label{fig:infographic}
\end{figure*}

For each treated and control county, we extract a time series of two-week cumulative counts of a given health outcome (e.g. mortality) for a 140-day period starting 129 days prior to the TC's first US approach and ending 11 days after. We use these time series to construct a (separate) panel data matrix for each TC and outcome, the general structure of which is illustrated in  Figure~\ref{fig:infographic}A. While the figure is intentionally left general, in our context only the final two-week window in the time series (the final column in the matrix) is considered to be the ``treatment period'' for treated counties. This corresponds to a two-week period beginning 2 days prior to the TC's arrival and ending 11 days after \citep{yan2020tropical}. For the predictive component of our model, we also obtain county-specific TC features (e.g., maximum sustained windspeed, duration of sustained wind speeds above 20 m/s) and socioeconomic and demographic characteristics for each TC and each exposed county.

\subsection{Approach}\label{ss:approach}

%We propose a Bayesian machine learning model to simultaneously estimate the health effects of historic TCs in each impacted county and to capture the relationship between those estimated health effects and the features of the TC and the impacted populations, in order to predict the excess health events expected for an impending TC with .
For each health outcome, we construct a model composed of (1) causal inference sub-models for each TC to estimate the excess health events attributable to it in each impacted county and (2) a predictive sub-model that relates these health effects to the TC and county features. We emphasize that each health outcome is modeled separately, with no transfer of information between the outcome-specific models. For the remainder of the section, we focus on the model for a single outcome. To emphasize the broader applicability of our approach, we present the methods using general notation, making connections back to our specific TC data structures for clarity.

\subsubsection{Causal inference sub-models}\label{s:s:s:causal}

In this section, we describe the models that will be used to estimate the excess health events attributable to historic TCs. These models are applied separately to the data for each TC, which is part of a larger modularized model fitting scheme described in Section~\ref{sss:modular}.

We denote the number of TCs in the study by $S$. In the causal inference sub-models, all data and parameters are storm-specific and should be indexed by an $s\in \left\lbrace 1, \dots, S \right\rbrace$. However, for clarity of presentation, we suppress these indices and introduce the causal inference concepts in the context of a single arbitrary TC. Let $i=1,...,N$ index the set of treated and control counties and $t=1,...,T$ index time windows, so that the panel data matrix (Figure~\ref{fig:infographic}A) has dimensions $N\times T$, with counties in the rows and times in the columns.

For a given TC, we assume that we have a set of ``treated'' and control counties, that all counties are untreated at $t=1$, and that once treatment begins for treated counties they remain treated through $t=T$. We let $W$ denote the set of indices of treated counties. Let $D_{it}$ be a binary indicator of treatment of county $i$ by the TC at time $t$. Let $T_0$ denote the common time period when treatment is initiated in treated counties, such that all counties are untreated prior to $T_0$, and treatment occurs for treated counties at times $T_0$ through $T$ (see Figure ~\ref{fig:infographic}A). While staggered treatment initiation times can be accommodated in this framework, we focus on a common treatment initiation time for clarity. Thus, $D_{it}=1$ if $\left\lbrace (i,t): i\in W, t \geq T_0\right\rbrace$, $D_{it}=0$ otherwise. Collectively, the set of all $t\in \left[ T_0, T\right]$ is referred to as the treatment period. %In our analyses, $T_0 = T$ because we choose not to include any two-week time periods after the TC exposure period $T_0$, but we intentionally leave the notation general. 

$Y_{it}$ is the observed number of health events for county $i$ at time $t$, and the panel data matrix of the outcomes is denoted $\mathbf{Y}$. We formalize our causal inference approach using Rubin's potential outcomes framework \citep{rubin1974estimating}, invoking assumptions given in Section~\ref{s:s:ident}. In short, in treated counties during treatment, we observe the outcome that occurs under treatment and we wish to compare it to an estimate of the ``counterfactual'' outcome that would have occurred in the absence of treatment. Formally, let $Y_{it} (0)$ be the potential outcome in county $i$ at time $t$ under control. For control counties at all times and for treated counties prior to the treatment period, $Y_{it} (0)=Y_{it}$. For treated counties during the treatment period, we instead observe the potential outcome under treatment, $Y_{it} (1)=Y_{it}$. The aim of the causal inference sub-model for each TC is to estimate the individual excess events (IEE), defined as
%$\theta_{i}=\frac{1}{T-T_0}\sum_{t\geq T_0} \left[ Y_{it} (1)-Y_{it} (0) \right]$  for $i\in W$. 
$\theta_{i}=\sum_{t\geq T_0} \left[ Y_{it} (1)-Y_{it} (0) \right]$  for $i\in W$. Here the word individual refers to individual units of analysis, in our case counties.
Because $Y_{it}(1)=Y_{it}$ for $\left\lbrace (i,t): i\in W, t \geq T_0\right\rbrace$, our aim is to estimate the counterfactual outcome, $Y_{it} (0)$. 

Both spatial and temporal confounding are possible in studies of the health impacts of TCs. For example, coastal TC-prone counties may have wealthier populations and wealth is associated with health. Alternatively, TCs may be more likely to occur under certain climate conditions which may independently affect health outcomes. With most observational study designs, causal inference analyses rely on the assumption of ignorable treatment assignment conditional on observed confounders (no unmeasured confounding). To flexibly address potentially unmeasured confounding, we conceptualize each TC as a quasi-experiment, i.e., a study design with nonrandomized treatment assignment but with pre- and post-treatment data available. In environmental health studies, quasi-experimental designs are the gold standard for assessing causality because certain types of unmeasured confounders can be controlled for by design \citep{dominici2017best}.

Classic methods such as difference-in-differences allow for control for time-invariant unmeasured confounders. Recent machine learning approaches such as matrix completion \citep{athey2018matrix} go further by allowing control for certain types of time-varying unmeasured confounders. This ability to adjust for time-varying unmeasured confounding is particularly critical in our TC application. Many potential confounders of TC health effects demonstrate complex seasonal patterns, e.g., employment \citep{krane1999cyclical}, use of homeless shelters \citep{colburn2017seasonal}, and infectious disease proliferation, but measurements of these variables are unavailable at the space-time resolution needed.
%as county-level potential confounders are not generally measured at the sub-year time scales that would be needed for traditional confounding adjustment in our context.
%In general, in order to draw causal conclusions from analyses of observational data, we must make the assumption that there are no unmeasured confounders. However, natural experiments provide an opportunity to adjust for many types of unmeasured confounders. To estimate the health impacts of a TC in each treated county, we propose an adaptation of the matrix completion (MC) approach proposed by Athey et al \citep{athey2018matrix} for conducting causal inference on natural experiments using panel data. MC is advantageous because it adjusts for unobserved time-varying confounders under mild assumptions.

% Confounding is possible in studies of the health impacts of TCs because, for example, coastal TC-prone counties may have wealthier populations and wealth is associated with health.

To estimate the health impacts of a TC in each treated county, we propose an adaptation of the matrix completion (MC) approach for conducting causal inference on natural experiments using panel data \citep{athey2018matrix,tanaka2019bayesian,pang2020bayesian}. MC is a machine learning technique for imputing missing values in a matrix, learning from patterns in observed entries in both the rows and columns. In our setting, the matrix with missing entries is the matrix of $Y_{it} (0)$ values, denoted by $\mathbf{Y(0)}$. $\mathbf{Y(0)}$ is structured just like the panel data matrix, with missing entries in positions corresponding to the treated counties during the treatment period (Figure~\ref{fig:infographic}A, blue elements missing). MC learns from space-time trends in the non-missing data, i.e., the outcomes for (1) control counties at all time periods and (2) treated counties prior to treatment, to impute the missing $Y_{it} (0)$. In this approach, the observed $Y_{it}(1)$ are treated as fixed and known and are entirely omitted from the MC model. In settings with normally distributed data, MC can be framed as a factorization of $\mathbf{Y(0)}$ (or of its expectation), as illustrated in Figure~\ref{fig:infographic}B. 
%For normally-distributed data, Athey et al specify the MC model as $Y_{it} (0)=\mathbf{U}_i^T \mathbf{V}_t+\epsilon_{ij}$ (equivalently, $\mathbf{Y(0)=U^T V+\epsilon}$; Figure~\ref{s:fig:mc}) \citep{athey2018matrix}. Here, $\mathbf{V}$ is a $K \times T$ matrix with columns $\mathbf{V}_t$ representing a small number of factors influencing the $Y_{it} (0)$ that vary over time but are common to all counties and $\mathbf{U}$ is a $K \times N$ matrix with columns $\mathbf{U}_i$ representing the county-specific effects of the $\mathbf{V}_t$ on $Y_{it} (0)$. Together, $\mathbf{V}$ and $\mathbf{U}$ provide a low-dimensional representation of the space-time trends in the $Y_{it} (0)$. For a treated county $i'$ at post-treatment time $t'$, its missing $Y_{i' t'} (0)$ is assumed to follow a distribution centered at $E\left[Y_{i't'}(0)\right]=\mathbf{\hat{U}}_{i'}^T \mathbf{\hat{V}}_{t'}$ (see Figure~\ref{s:fig:mc}).

% Here, $\mathbf{V}_t$ is a $K$-length vector of time-specific latent factors, $\mathbf{U}_i$ is a $K$-length vector of unknown county-specific factor loadings, and $K$ is an unknown scalar with $K<<min⁡(N,T$). The $\mathbf{V}_t$ represent a small number of factors influencing the $Y_{it} (0)$ that vary over time but are common to all counties, and the $\mathbf{U}_i$ can be viewed as coefficients representing the county-specific effects of the $\mathbf{V}_t$ on $Y_{it} (0)$. 

Because our outcomes are counts, we generalize the MC approach for causal inference to allow for count data likelihoods. MC models for count data were developed in other contexts \citep{gopalan2014content}, but do not follow epidemiologic conventions for modeling count data. We instead propose the following MC model for count data using a log link:
\begin{equation}\label{eq:bmc}
    \text{log}(E\left[Y_{it} (0)\right])=\alpha+\gamma_i+\psi_t+\mathbf{U}_i^T \mathbf{V}_t+log(p_{it})
\end{equation}
$\alpha$ is a global intercept, $\gamma_i$ are county-specific deviations from the global intercept, and $\psi_t$ are time-specific deviations. $\mathbf{V}_t$ is a $K$-length ($K<<min(N,T)$, unknown) vector of unobserved factors influencing the $Y_{it} (0)$ that vary over time but are common to all counties and $\mathbf{U}_i$ is a $K$-length vector of the unobserved county-specific effects of the $\mathbf{V}_t$ on $Y_{it} (0)$. Together, the $\mathbf{V}_t$ and $\mathbf{U}_i$ provide a low-dimensional representation of the space-time trends in the $Y_{it} (0)$ (see Figure~\ref{fig:infographic}B for an illustration in the case of normally-distributed outcomes). $p_{it}$ is a scalar population size offset, to allow for rate outcomes. %In our case, the offset used for the cardiovascular, chronic obstructive pulmonary disorder, and respiratory hospitalization models is the county's Medicare fee-for-service population, and that for the mortality count model is the county's entire Medicare population. 
We pre-specify $K$ based on exploratory principal component analyses and fit the MC models using a negative binomial likelihood and uninformative prior distributions, collecting MCMC samples using the rstan \citep{rstan} software package. Explicit modeling details are given in Section~\ref{s:s:bayes_model}. For a treated county $i$ at post-treatment time $t$, we use the above model to collect $M$ MCMC samples from the posterior predictive distributions of the missing counterfactuals, denoted $Y_{it}^{(m)}(0)$ for $m=\left\lbrace 1,...,M\right\rbrace$, and use those to construct $M$ posterior samples of the IEE, as $\theta_i^{(m)}=\sum_{t\geq T_0} \left\lbrace Y_{it}-Y_{it}^{(m)} (0) \right\rbrace$ for $i\in W$.

The formal causal identifying assumptions for this model, originally specified in  \citep{pang2020bayesian} are provided in Section~\ref{s:s:ident}. Under these assumptions,  the $\mathbf{U}_i^T \mathbf{V}_t$ should capture all space-time trends in the $Y_{it}(0)$, including trends induced by time-varying confounders. Thus the resulting IEE can be identified, assuming trends in confounders do not change differentially in treated units (relative to controls) post-treatment. 

%Additional advantages of the Bayesian approach include the stability imparted by Bayesian shrinkage properties and avoidance of asymptotic approximations. 

In practice, both the excess number of events and the excess rate of events (per unit population) are of interest for understanding the epidemiology of extreme weather events. Thus we define the individual excess rate as $\theta^*_{i}=100000\times (\theta_{i}/p_{iT})$. We also define TC-specific excess events as the cumulative excess events across all counties impacted by a TC, and TC-specific excess rate as the excess rate across all impacted counties. To compare with existing literature and evaluate overall health burdens, we also wish to summarize the estimated health effects across our entire study. To this end, we define the total excess events (TEE) for the full study to be the cumulative TC-attributable excess events summed over all TCs and counties, and the average excess rate (AER) to be the average of the excess rates across all county-level TC exposures in the study. Formal definitions of each estimand are given in Table~\ref{s:tab:estimands}.
Posterior samples of these quantities can be constructed through simple transformations of the $\theta_i^{(m)}$.

\subsubsection{Bayesian modularization}\label{sss:modular}

%One advantage of Bayesian approaches for complex models is the ability to propagate uncertainties between model components.
In a classic Bayesian framework, a full likelihood is specified for the data, and the model components are fit jointly, permitting unrestricted information flow. However, in many real-world contexts, there is a need to propagate uncertainty between model components without allowing information to flow bi-directionally between all model components. 
%One notable example is Bayesian propensity score methods in causal inference, where dominant causal inference philosophies suggest that uncertainties from the propensity score model should be propagated to the outcome model, but the outcome model should not be allowed to influence the posterior distribution of the propensity score\citep{mccandless2010cutting,zigler2013model}.
This may be due to philosophical considerations, as in the case of Bayesian propensity score methods \citep{mccandless2010cutting,zigler2013model}, or practical considerations, as complex models fit jointly may suffer from poor mixing or require prohibitive computation times. These concerns have given rise to a literature on Bayesian modularization, in which information flows between certain sub-models weakly or not at all  \citep{liu2009modularization,lunn2009combining,plummer2015cuts,jacob2017better}. This is often achieved by ignoring some components of the joint likelihood.

We modularize our models in a manner that prevents information flow between the causal inference sub-models for each TC (as described above), yet allows information to flow uni-directionally from the causal models into the predictive model. This permits uncertainty in the TC health effect estimates to be accounted for when fitting the predictive model, but does not allow the predictive model to inform the causal effect estimates. Explicit details are provided in Section~\ref{s:s:modular}. This modularization approach is motivated by both philosophy and computational feasibility. Primarily, we wish to prevent information from the predictive model from influencing the causal models, which would obscure the identifying assumptions needed to obtain causal effect estimates. Because our causal modeling approach is computationally intensive and involves many unidentifiable parameters, the modularization approach is also more practical, as it improves mixing and reduces computation time by enabling parallel model-fitting across TCs. 

%brooke.anderson: One of the really cool things about this model is how it tackles something hard to do in terms of predictive model building. We have an outcome (excess mortality, excess hospitalizations) that we *cannot* measure we certainty. There is inherent variance/uncertainty in our estimate of this for each storm. However, from a practical/public health point of view, it's *very* important (as demonstrated by Maria). There are *not* loads of good methods to build a predictive model when you've got an outcome that's measured/estimated with uncertainty, right? If we're going to the Statistics section of PNAS (which we certainly want to, I think), *that's* the part that will get statisticians/machine learning folks excited, and that will be something that could be applied in a lot of other contexts outside of epidemiology/tropical cyclones. We're hinting at this point, in terms of the novelty of our study, here, but I think that we'll get more readers excited if we really clarify this point and have it as a running thread throughout the paper---to build such a model for TCs, we must have a way to build a predictive model when the target outcome is estimated, not measured with certainty, and we bring together cutting edge approaches from causal inference and machine learning to do that.

\subsubsection{Predictive sub-model}

We develop a predictive model for each health outcome that captures the relationship between the county-specific TC health effects and the features of the TC and county (i.e., characterizing how such features modify TC health impacts). For clarity in this section, we re-introduce the storm-specific indices, but continue to focus on a single outcome-specific model. We let $\theta_{si}^{*(m)}$ be the individual excess rate posterior sample $\theta_i^{*(m)}$ for TC $s$, and $W_{s}$ be the set of indices of treated counties $W$ for TC $s$. Then, for a single fixed posterior draw $m$, we collect a posterior sample of the parameters $\beta$ from the (outcome-specific) predictive model:
\[
\{\theta_{si}^{*(m)}=g(\mathbf{X}_{si};\mathbf{\beta}) \; | \; s \in \{1, \dots S\}, i \in W_{s} \}
\]
where $\mathbf{X}_{si}$ is a vector of predictors, i.e., modifiers, of the county-specific TC effects, and $g()$ is an unspecified function parameterized by a vector of global parameters $\mathbf{\beta}$. In practice, $g()$ could take the form of any Bayesian predictive model. We recommend selecting $g$() based on cross-validation performance. We repeat this sampling with each $\theta_{si}^{*(m)}$ to obtain posterior samples $\left\lbrace \beta^{(1)},...,\beta^{(M)}\right\rbrace$ (Figure~\ref{fig:infographic}C).

\subsubsection{Prediction for future TCs}
Using the posterior samples $\mathbf{\beta}^{(m)}$, we can draw corresponding posterior predictive samples of the health effect, $\theta_{new}^{*(m)}$, for any set of predictor values $\mathbf{X}_{new}$. To use the model for county-level prediction of the health impacts of a specific approaching TC, $\mathbf{X}_{new}$ could be defined as the forecasted meteorological characteristics of the TC and socioeconomic and demographic characteristics of each county on its expected path. The predicted health impacts and uncertainties for each county can be used to identify counties at highest health risk. Alternatively, to create a long-term TC health risk profile for a county, many different $\mathbf{X}_{new}$ vectors could be created using the meteorological characteristics of a collection of hypothetical, representative TC exposures, as well as the socioeconomic and demographic characteristics of the county. The resulting set of predictions can be summarized to give insight into future TC health risks the community may face, in both expected and extreme TC scenarios.

\begin{figure*}%[\sidecaptionrelwidth][t]
\centering
\includegraphics[scale=.55]{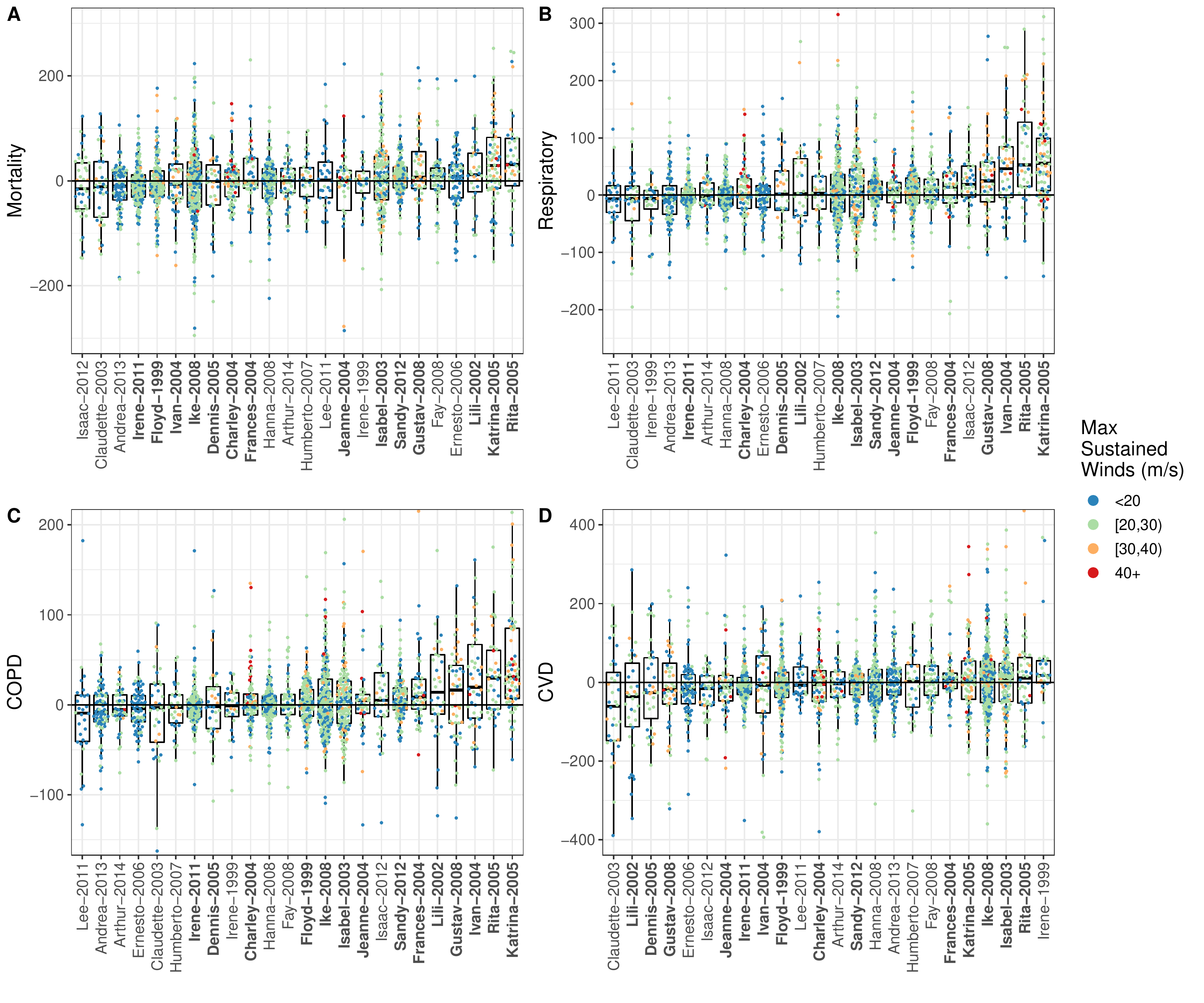}
\caption{County-level excess rate estimates for mortality (A), respiratory hospitalizations (B), COPD hospitalizations (C), and CVD hospitalizations (D) for TCs that impacted $>25$ counties. The county excess rate is the estimated rate of excess events (per 100,000) in the county due to the TC. Distant outliers are cropped out for readability. Bolded TC labels indicate storm names that were subsequently retired-- retirement occurs when a TC is so destructive that re-using the name is considered to be insensitive \citep{nhc2020}.}
\label{fig:rate_beeswarm}
\end{figure*}

\section{Results}\label{s:results}

\subsection{Causal analysis}
%A total of 57 Atlantic basin TCs exposed at least one US county during the period 1999-2015.
%to a sustained windspeed of at least 17.5m/s.
53 TCs and 2,135 corresponding county-level TC exposures occurring during the period 1999-2015 are included in our analysis (see inclusion criteria in Section~\ref{s:s:paneldata}). In Table~\ref{s:tab:storm_feat}, we provide the name and year of each TC included in our study, the number of treated and control counties used in its causal model, and the rate of each health outcome among the treated and controls during the 140-day period surrounding the storm. Figure~\ref{s:fig:tc_hits} maps the number of TC exposures by county. Coastal counties in the Carolinas and the Gulf Coast region are repeatedly exposed, with some receiving as many as 15 TC exposures during our 17-year study period. For a discussion of the possible impacts of TC-related population displacement on our analyses, see Section~\ref{s:s:pop_displace}.

We apply the MC models for each TC and health outcome with $K=4$ factors. $K=4$ was chosen because exploratory principal component analyses revealed that 4 factors explained around 70\% of the variance in the $\mathbf{Y(0)}$ matrices (Section~\ref{s:s:select_k}). This selection allows for preservation of critical variance without overfitting. %Visual evaluation of model fit also indicated that $K=4$ produced $Y_{it}(0)$ predictions that captured the trends in the observed $Y_{it}(0)$ values in treated counties prior to treatment.
We run the causal models using two separate MCMC chains, collecting 1000 post-burn-in samples from each chain. Traceplots of the $Y_{it}^{(m)}(0)$ indicated convergence.

% known county population size at time $T$.

% $$\frac{1}{\sum_{i=1}^{|W_s|}p_{iT}} \sum_{i=1}^{|W_s|} \theta_{si}$$

\begin{figure*}
\centering
\includegraphics[scale=.55]{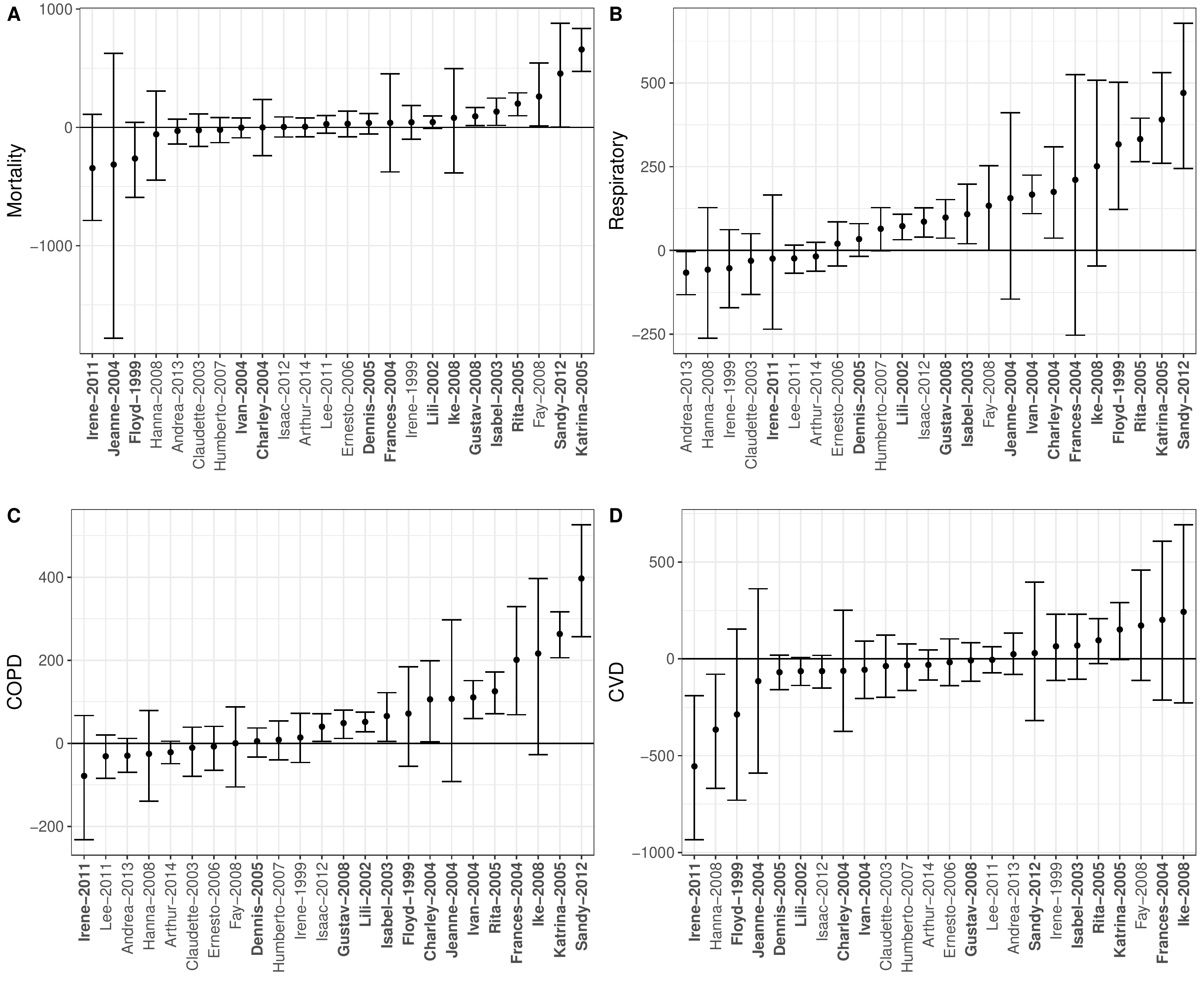}
\caption{TC-specific excess events estimates and 95\% predictive intervals for mortality (A), respiratory hospitalizations (B), COPD hospitalizations (C), and CVD hospitalizations (D) for TCs that impacted $>25$ counties. The The TC-specific excess events is the sum of excess event estimates across all counties impacted by the TC. Bolded TC labels indicate storm names that were subsequently retired-- retirement occurs when a TC is so destructive that re-using the name is considered to be insensitive \citep{nhc2020}.}
\label{fig:tee}
\end{figure*}

\subsubsection{TC- and county-level estimated health effects}

Recall that our analysis defines the treatment period as only the final two-week time window (beginning two days prior to the storm's first approach and ending 11 days after). Thus, the IEE for county $i$ exposed to TC $s$, can be expressed simply as $\theta_{si}= Y_{iT} (1)-Y_{iT} (0)$, i.e., the excess health events attributable to the TC at time $T$. For each of the four health outcomes, we have generated posterior samples of the IEE for each county impacted by each TC. We use these to construct posterior samples of the excess rates and the summary quantities described in Section~\ref{s:s:s:causal}.
%, i.e., we compute the sum/average of ITTs corresponding to each posterior sample.
Hereafter, we refer to the posterior means for each parameter as the ``estimates'' from our models. Figure~\ref{fig:rate_beeswarm} shows the county-level excess rate estimates, grouped by TC, for all TCs that impacted $>25$ counties. Figure~\ref{s:fig:att} gives the TC-specific excess rate estimates. Figure~\ref{fig:tee} displays the TC-specific excess event estimates and 95\% credible intervals, and the county-level excess event estimates (IEE) are shown in Figure~\ref{s:fig:excess_beeswarm}. These results illustrate the heterogeneity in TC health effects across counties and across storms.

We find that on average a county's mortality rate increases slightly, though not significantly, during the 2-week treatment period, compared to the mortality rate expected in the absence of TC (AER: 2.58, 95\% CI [-1.69, 6.56]; TEE: 1228.86, 95\% CI [-608.20, 2731.07]). TC exposures cause larger and significant increases, on average, in respiratory hospitalizations (AER: 8.58 [4.34, 11.86]; TEE: 2926.18 [1808.97, 3940.02]) and COPD hospitalizations (AER: 4.57 [2.13, 6.79]; TEE: 1532.80 [969.95, 2106.10]). For each of these outcomes, we note that Hurricanes Katrina and Rita, which impacted largely overlapping sets of counties in the same year (2005), produced some of the largest adverse impacts (on both the excess event and the rate scale). We find that Hurricane Sandy caused huge increases in these outcomes specifically on the excess events scale, which is likely attributable to its impacts on the densely populated New York City area. Moreover, for each of these outcomes, Figures~\ref{fig:rate_beeswarm} and ~\ref{s:fig:excess_beeswarm} suggest that counties experiencing higher TC windspeeds may be at increased risk. 

For CVD, we find that hospitalization rates on average decrease during the 2-week period surrounding TC exposure (AER: -5.01 [-9.87, -0.30]; TEE: -977.99 [-2246.53, 222.10]). A previous study found decreases in CVD hospitalizations on the day of the storm but increases 2-3 days later \citep{yan2020tropical}.
%We also note from Figure~\ref{fig:rate_beeswarm} that there is no apparent association between TC windspeeds and the change in CVD hospitalization rates.
This finding is likely attributable in part to the fact that our hospitalization metric captures all inpatient hospitalizations, including planned procedures. The danger associated with venturing out during or immediately after a TC may motivate people to cancel planned procedures or treatment for chronic disease, so, while emergency CVD hospitalizations may have increased during the treatment period, the decrease in non-emergency CVD hospitalizations likely explains the overall reduction in CVD hospitalizations during the TC period.

\subsection{Predictive analysis}

The full set of candidate predictors is given in Section~\ref{s:s:features}. We conduct predictive model selection using cross-validation as described in Section~\ref{s:s:pred_mod_select}. The selected predictive model (common across health outcomes) is a Bayesian linear model with a spline on windspeed and year, with the remaining TC-related and socioeconomic/demographic predictors included as linear terms. For interpretability, we also provide results from a Bayesian linear regression model without the windspeed spline.

We fit the full modularized Bayesian models and obtain 1000 post-burn-in samples of the predictive model parameters. Tables~\ref{s:tab:pred_model_spl} and~\ref{s:tab:pred_model_lin} give the posterior means and 95\% credible intervals for the predictive model coefficients. To illustrate the importance of propagating uncertainties from the causal to predictive modules of our model, we also overlay our predictive model estimates and 95\% CIs with those obtained by implementing the causal and predictive models separately, i.e., without propagating uncertainty (Figure~\ref{s:fig:propagate}).

\begin{figure}[h]
\centering
\includegraphics[scale=.7]{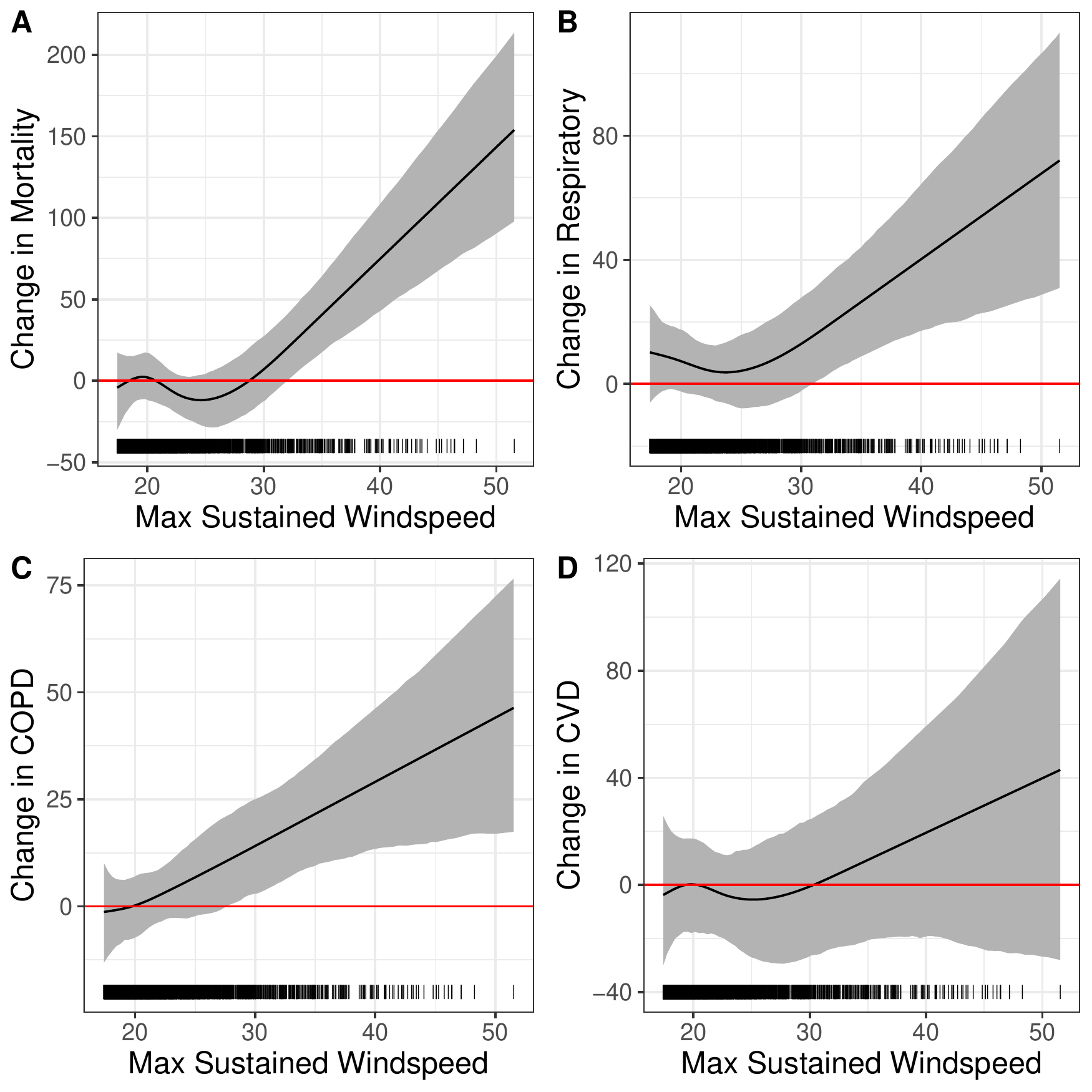}
\caption{Relationship between maximum sustained windspeed and county-level excess rates per 100,000 of mortality (A), respiratory hospitalizations (B), COPD hospitalizations (C), and CVD hospitalizations (D).}
\label{s:fig:ws_spline}
\end{figure}

In Figure~\ref{s:fig:ws_spline}, we show the windspeed splines and 95\% credible intervals for each outcome. For each outcome, Table~\ref{s:tab:pred_model_lin} indicates that maximum sustained windspeed has the strongest association with health impacts, among the predictors considered. The splines illustrate that, as windspeeds increase beyond 30m/s, we observe a sharp increase in TC-attributable mortality and respiratory and COPD hospitalizations. While we generally find a similar trend for windspeed and CVD hospitalizations, the relationship is weaker and more variable.

We also find that TC-attributable respiratory hospitalizations are positively associated with the duration of sustained windspeeds above 20m/s (Table~\ref{s:tab:pred_model_spl}). For respiratory and COPD hospitalizations, we observe a negative association with total number of TC exposures during the study period, a proxy for TC exposure propensity (Table~\ref{s:tab:pred_model_spl}). This suggests that communities that are frequently hit may adapt in ways that decrease respiratory health impacts (e.g., bury power lines to decrease power outages, thereby decreasing risk for those dependent on electric-powered respiratory devices). Although few strong associations are detected for the county socioeconomic and demographic features, we find that predominately White communities tend to experience fewer TC-attributable COPD hospitalizations (Table~\ref{s:tab:pred_model_spl}).

\subsection{Sensitivity analyses}

We conduct a range of sensitivity analyses for both the causal and predictive components of our model (Section~\ref{s:s:sensitivity}). For the causal models, we evaluate sensitivity to our definition of TC exposure and to specification of $K$. Because we have cumulative TC precipitation data only for years prior to 2012, we fit predictive models restricted to years 2011 and earlier and include a restricted cubic spline on cumulative precipitation as a predictor. In short, we find that our causal models are robust to these specifications and that precipitation is weakly, if at all, associated with the acute health impacts of TCs after adjusting for other factors.

\section{Discussion}\label{s:discuss}

We have proposed and implemented an integrated causal and predictive modeling framework for systematically characterizing and predicting the health impacts of TCs in the US, in order to inform pre-storm strategic preparedness efforts. This work offers several contributions to the existing literature on TC epidemiology. First, we have used a standardized causal inference approach to estimate county-level TC-attributable excess mortality and excess respiratory, COPD, and CVD hospitalizations (with uncertainties) in the Medicare population for nearly all Atlantic-basin TCs 1999-2015. These excess event estimates provide a more complete picture of TC health burdens than post-storm surveillance efforts or single-storm studies. We have also found that, controlling for a number of demographic and meteorological predictors, the maximum sustained windspeed experienced by a county is a strong predictor of its TC-attributable increases in adverse health events, potentially providing insight into strategies to minimize future TC health burdens. Our predictive models may also be useful for identifying specific communities facing the highest risk from future TC, which is critical to avert the most severe health consequences. Finally, this modeling approach can be used analogously in the context of other extreme weather and climate events, including heat waves, droughts, floods, and wildfire smoke exposures.

More detailed data on the multi-dimensional TC exposures and pre-TC preparatory measures would improve the predictive ability of our models and provide greater insight into how to minimize TC health burdens. For instance, flooding is a common and often devastating TC-related exposure. While a county-level binary indicator of TC flooding is available \citep{anderson2020assessing}, this is insufficient for understanding the impact of floods, which tend to be highly complex and localized. Additionally, mandatory pre-storm evacuation orders may be a critically influential factor in the health impacts of a TC; however, to our knowledge, evacuation data have never been systematically compiled on a multi-storm scale. To minimize the health threats presented by climate and weather disasters, we must continue to collect, compile, and analyze more and higher quality data on these events.

%%TC:ignore
\section{Acknowledgements}

The authors gratefully acknowledge funding from NIH grants R01HD092580, R01AG060232-01A1, R01 ES028805, R01 ES030616, P30 ES009089, R00 ES022631 and NSF grants NSF 1940141 and NSF 1331399.

\clearpage

% Bibliography
\bibliographystyle{chicago}
\bibliography{hurricanes}

\clearpage

\appendix
\counterwithin{figure}{section}
\counterwithin{table}{section}

\renewcommand{\thesection}{S}

\section{Supplementary Information: Integrated causal-predictive machine learning models for tropical cyclone epidemiology}

\subsection{Data}\label{s:s:data}
\subsubsection{Health outcomes}\label{s:s:outcomes}
This study focuses on the following adverse health events: all-cause mortality, respiratory hospitalizations, chronic obstructive pulmonary disorder (COPD) hospitalizations, and cardiovascular disease (CVD) hospitalizations (ICD-9/10 codes given in Table~\ref{s:tab:icd}). These outcomes were selected based on evidence from previous TC research and a biologically plausible link to TC exposures \citepSM{yan2020tropical}. Each individual's hospitalizations and death are assigned to their county of residence (not the county of hospitalization). We construct daily counts of CVD, respiratory, and COPD hospitalizations in the Medicare fee-for-service population for each county in the eastern US for the period 1999-2015. For this same set of counties, we also utilize daily county-level mortality counts in the entire Medicare population (not restricted to fee-for-service). When population denominators are used for the creation of incidence rates, the total Medicare population size is used for mortality outcomes, while the Medicare fee-for-service population size is used for hospitalization outcomes.

% latex table generated in R 3.6.3 by xtable 1.8-4 package
% Tue Aug 11 18:25:15 2020
\begin{table}[ht]
\centering
\caption{ICD-9/10 codes used to define each hospitalization type.}
\centerline{
\begin{tabular}{p{4cm}p{3cm}p{3cm}}
  \hline
Hospitalization Type & ICD-9 & ICD-10 \\ 
  \hline
Cardiovascular Diseases & 390.xx--398.xx,
              401.xx--405.xx,
              410.xx--414.xx,
              415.xx--417.xx,
              420.xx--429.xx & I00--I52\\
              & &  \\
Respiratory Diseases & 464.xx--466.xx,
               480.xx--487.xx,
               490.xx--492.xx,
               494.xx--496.xx & J04--J06,
                J20--J21,
                J09--J18,
                J40--J44,
                J47,
                J67\\
                & &  \\
Chronic Obstructive Pulmonary Disease & 490.xx--492.xx, 494.xx--496.xx & J40, J410, J411, J449, J441, J440, J418, J42, J439, J479, J471, J670--J679\\
   \hline
\end{tabular}}
\label{s:tab:icd}
\end{table}

\subsubsection{TC exposures}\label{s:s:exposure}
\begin{figure}[h!]
\centering
\includegraphics[scale=.6]{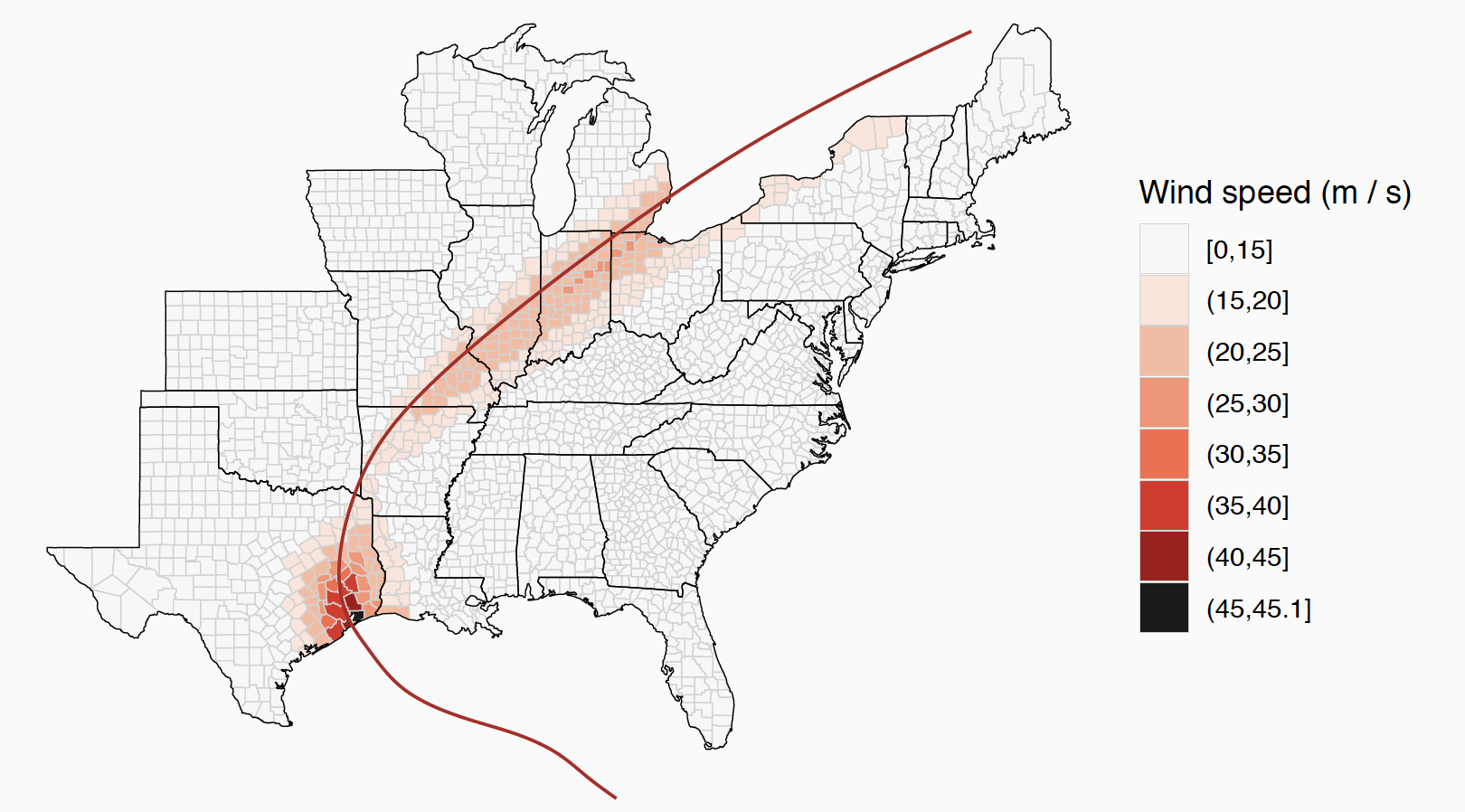}
\caption{Illustration of TC track and wind exposure measure for Hurricane Ike, 2008.}
\label{s:fig:exposure}
\end{figure}

To characterize county-level TC exposure, we leverage an open source data platform containing temporally-detailed track and feature data for each Atlantic-basin TC during the period 1999-2015 that came within 250km of at least one eastern US county. This data platform, which is made available through the \texttt{hurricaneexposuredata} R package, has been fully described elsewhere \citepSM{anderson2017Rdata,anderson2017Rpack,anderson2020assessing}. Briefly, the TC tracks are obtained from the US National Hurricane Center's ``Best Tracks'' dataset, which records the storm's position every 6 hours, and the storm's position is interpolated to 15-minute intervals. Wind fields are then modeled at 15-minute intervals, and the resulting values are used to estimate the storm's maximum sustained windspeed at the population centroid of each US county (see Figure ~\ref{s:fig:exposure}). Cumulative precipitation amounts in each county associated with the TC are estimated by summing rainfall amounts from the North American Land Data Assimilation System Phase 2 (NLDAS-2) re-analysis dataset \citepSM{rui2014readme} over a 4-day window beginning two days prior to the storm's closest approach to the county (these precipitation data are currently available only for years 1999-2011). The county-level wind and precipitation exposure metrics have undergone validation for use in epidemiologic studies \citepSM{anderson2020assessing}.

While the continuous windspeed and rainfall metrics are employed in the predictive component of the model, for the causal inference component we must define a binary metric of TC exposure, which we refer to as ``treatment'' for consistency with the causal inference literature. Counties exposed to the TC are referred to as treated counties and those not exposed as control counties. Because previous epidemiologic research has suggested that TC windspeed is the feature most associated with acute health impacts \citepSM{yan2020tropical}, we use a maximum sustained windspeed threshold to classify counties as treated or control. Moreover, because TC windfields are large and relatively homogeneous over small areas, their use to define TC exposure reduces within-county variation and the potential for exposure misclassification. We classify a county as treated for a given storm if it experienced sustained gale-force or higher sustained wind speeds ($\geq$ 17.4  meters/second); otherwise, the county is classified as a control \citepSM{yan2020tropical}. This threshold is consistent with the outer limit wind threshold used in the US National Hurricane Center's wind radii product for characterizing tropical cyclone size. %We also conduct sensitivity analyses using a sustained windspeed threshold of 32.5 meters/second, which is commonly used to classify a TC as hurricane-strength.

\subsubsection{Exposure and health data linkage}\label{s:s:paneldata}
Our analytic approach requires defining a study period for each TC, composed of a substantial time span prior to the storm (used to establish baseline trends in health outcomes in treated and control counties) as well as the period during and immediately after the storm (to estimate acute storm effects). For each TC, we define its study period as beginning 129 days prior to the TC's first US approach and ending 11 days after. The TC's first approach is the earliest date that the TC makes its closest approach to an exposed US county/counties, which can be obtained from the \texttt{hurricaneexposuredata} package \citepSM{anderson2020assessing}. In practice, the approximately 18 weeks of outcome data prior to the TC exposure for each county is aggregated into nine two-week counts of the outcome. Two-week cumulative counts are used because narrower time intervals leads to small counts and instability in time series for many counties. We choose to employ nine of these two-week counts because (1) this provides enough time points to reveal relationships in baseline trends in the health outcomes in treated vs. control counties while (2) the time series covers a limited enough period that relationships between the time trends in treated and control counties would be expected to remain stable. The use of longer time series may introduce unnecessary noise into the models by capturing irrelevant long-term changes in relationships in baseline health across counties.

After designating each county in the study as treated or control for each TC as described above, we select a set of analytic treated and control counties for a given TC as follows. Any county (treated or control) is automatically excluded from the analytic set for a given TC if it has Medicare fee-for-service population size less than 100, or if it experiences 5 or fewer of any of the health events under study during the TC study period. Further, among the control counties, we select as analytic controls only those that fall within 150 miles of at least one treated county (distance computed between county centroids). Hereafter, when we reference the treated/control counties for a given TC, we are referring to the analytic treated/controls as defined here. We exclude TCs from the analysis if they have no valid treated counties, total sample size (treated + control counties) less than 20, or less than 5 control counties.

For each TC and each health outcome, we create a matrix of panel data, and these are central to the causal inference component of our model. The panel data matrix is composed of rows corresponding to each county in the analytic set for the given TC. Each row contains a time series of counts of the outcome in that county, for each two-week interval during the TC study period. This results in 10 two-week intervals, i.e., 10 columns in the panel data matrix. We then define each time period for each county as a control period or treatment period. For the control counties, all time periods represented in the panel data matrix are control periods. For treated counties, we consider the final two-week period, beginning 2 days before and ending 11 days after the TC's first approach, to be the treatment period. 

\subsubsection{County and storm-specific features}\label{s:s:features}

Our predictive model will be used to capture the association between a TC's health impacts and the TC's meteorological features and the socioeconomic and demographic characteristics of the affected communities. To this end, for the predictive component of our model, we utilize the following TC features, which are described in more detail by  \citeSM{anderson2020assessing}, for each TC-county pair (abbreviated variable name in parentheses):
modeled maximum sustained wind speed at the population centroid of the county during the TC (vmax\_sust),
duration of sustained wind speeds above 20 m/s at the population centroid of the county during the TC (sust\_dur),
year the TC occurred (year),
and total number of TC exposures experienced by the county during 1999-2015 (exposure), which is a proxy for TC exposure propensity. For years 1999-2011, cumulative precipitation  (precip) over a 4-day window, beginning two days prior to the TC's closest approach, is also available from the \texttt{hurricaneexposuredata} package. Because TC precipitation data are not available for our full study period, we use it only for sensitivity analyses for the predictive models.

We also consider the following spatial, demographic, and socioeconomic features of the county:
percent of residents in poverty (poverty),
percent of residents self-identifying as White (white\_pct),
%percent of Medicare-qualifying residents in poverty (poverty\_mcare),
percent of homes that are owner-occupied (owner\_occupied),
percent of residents age 65+ (age\_pct\_65\_plus),
%median age of residents (median\_age),
population density (population\_density),
median house value (median\_house\_value),
percent of residents without a high school degree (no\_grad),
%percent of Medicare-qualifying residents without a high school degree (no\_grad\_mcare),
a binary indicator of whether the county is located along the coastline (cc1), and
the state the county is located in (state). These variables are extracted from US Census Bureau products. In particular we rely on the decennial census and American Community Survey (ACS). We align the demographic and socioeconomic features for each TC-county pair with the year of TC exposure. For years when census or ACS values are available, we use them directly, and in years not covered by the ACS or decennial census, we rely on interpolated values.

\clearpage

\subsection{Bayesian MC model fitting details}\label{s:s:bayes_model}

Numerous Bayesian approaches to MC model fitting have been proposed  \citepSM{lim2007variational,salakhutdinov2008bayesian,gopalan2014content,yang2018fast}, and have been studied in the context of MC for causal inference  \citepSM{tanaka2019bayesian,pang2020bayesian}. Without further restrictions on the MC model, the elements of $\mathbf{U}$ and $\mathbf{V}$ are not uniquely identified (only identified up to an orthogonal rotation). However, in a Bayesian framework, the presence of unidentifiable parameters in the model does not compromise the estimation of identifiable parameters. Here, our interest lies in estimating $E\left[Y_{it}(0)\right]$ for missing entries, which is identifiable. Therefore, we allow $\mathbf{U}$ and $\mathbf{V}$ to remain unidentifiable and utilize instead the posterior distribution of $E\left[Y_{it}(0)\right]$ and the posterior predictive distribution of $Y_{it}(0)$ for inference.

Bayesian estimation proceeds by first specifying a data likelihood and prior distributions for the parameters. We specify the following likelihood: 
$$P(Y(0)|\mathbf{U,V},\eta)=\prod_i \prod_t \left[f(Y_{it} (0)|\mathbf{U}_i,\mathbf{V}_t, \eta)\right]^{1-D_{it}}$$
%$$P(Y(0)|\mathbf{U,V},\eta)=\prod_{i \not \in W} \prod_{t \ge T_0} \left[f(Y_{it} (0)|\mathbf{U}_i,\mathbf{V}_t, \eta)\right]$$
where $f$ is a negative binomial probability mass function with mean given by Equation~\ref{eq:bmc}, and a common scale parameter $\eta$. We fit the MC models using rstan \citepSM{rstan}, and to increase mixing and facilitate convergence we use the default flat prior distributions in Stan. Letting $U_{ki}$ be the element in the $k^{th}$ row and $i^{th}$ column of $\mathbf{U}$ and $V_{kt}$ be the element in the $k^{th}$ row and $t^{th}$ column of $\mathbf{V}$, then prior distributions are specified as follows:
\begin{equation}
\begin{split}
U_{ki}&\sim Uniform(-\infty,\infty)\\
V_{kt}&\sim Uniform(-\infty,\infty)\\
\eta &\sim Uniform(0,\infty)
\end{split}
\end{equation}
For each parameter, we sample from its distribution conditional on all of the other parameters in the same causal sub-model using MCMC sampling. %We then collect $M$ MCMC samples from the posterior predictive distributions of the missing counterfactuals, denoted $Y_{it}^{(m)}(0)$ for $m=\left\lbrace 1,...,M\right\rbrace$, and then use them to construct $M$ posterior samples of the ITT, as $\theta_i^{(m)}=\frac{1}{T-T_0}\sum_{t\geq T_0} \left\lbrace Y_{it}-Y_{it}^{(m)} (0) \right\rbrace$ for $i\in W$.

\clearpage

\subsection{Causal identifying assumptions}\label{s:s:ident}

In the most general setting, the potential outcomes for our panel data would be defined as $\mathbf{Y(d)}$, a matrix containing all outcomes when treatment statuses for all units at all time points are given by the matrix $\textbf{d}$. In this general case, each unit's potential outcomes depend on the treatment status of all units at all time points. In order to define the potential outcomes more concisely as we have in the paper, we invoke the classic Stable Unit Treatment Value Assumption used in causal inference \citepSM{rubin1980randomization}, which states that 1) there is only one version of treatment and that 2) one unit's outcome is unaffected by the treatment status of other units. If SUTVA is met, then we can define a unit's potential outcomes at time $t$ as $Y_{it} (\mathbf{d}_i)$, where $\mathbf{d}_i$ is a binary $T$-length vector of the treatment status of unit $i$ at times $1,...,T$.

In order to define a unit's potential outcomes at time $t$ as a function of only its treatment status at time $t$, as we have done in the main text, we are implicitly making the standard assumption that future treatment cannot impact past outcomes and making an additional assumption about the impact of treatment histories. 
%That is, a unit's potential outcomes at any given time point, $t$, do not depend on its full historic treatment status vector but rather only the unit's time of initial treatment adoption.
That is, we assume that a unit's full historic treatment status vector is a deterministic function of the unit's time of initial treatment adoption, so a unit's potential outcomes can be expressed as functions of only the unit's time of initial treatment adoption, rather than functions of its full historic treatment status vector. The panel data construction described in the main text is designed to satisfy this assumption, by requiring that treated units initiate treatment at some time $T_0>1$ and remain treated through time $T$. In general, we can allow for varying initiation times by unit, denoted by $T_{0i}$. In this setting, the treatment assignment vector $\mathbf{d}_i$ is determined entirely by treatment initiation time $T_{0i}$, such that we can write the observed treatment vector as $\mathbf{d}_i(T_{0i})$. Under this assumption, we can write $Y_{it}(\mathbf{d}_i=\mathbf{0})=Y_{it}(0)$, since treatment history is fully determined for a unit that has not yet initiated treatment. For treated units, because $Y_{it}(\mathbf{d}_i(T_{0i}))$ is observed when $t\geq T_{0i}$, it can be treated as a known quantity and simplified to $Y_{it}(1)$. Then for $i\in W,\; t\geq T_{0i}$ we can define the IEE as $\theta_{i}=\sum_{t\geq T_0} \left[ Y_{it} (1)-Y_{it} (0) \right]$.

We posit 4 additional assumptions that are needed to identify the causal effects of a TC using the proposed matrix completion approach.

\vspace{.5cm}

\textit{Assumption 1: Causal consistency.} This assumption states that the observed outcome is equal to the potential outcome under the observed treatment level, i.e.,
\[
Y_{it}(d)=Y_{it} \; \text{ if } \; D_{it}=d
\]

\vspace{.5cm}

\textit{Assumption 2: Latent ignorability.} Let $\mathbf{Y}_i(\mathbf{0})$ be the vector of outcomes under control for unit $i$ and $\mathbf{Y}_i(\mathbf{0})^{obs}$ be the observed elements of $\mathbf{Y}_i(\mathbf{0})$. This assumption states that conditional on observed covariates used in the causal model, $\mathbf{Z}_i$ (if any are used), the latent variables, $\mathbf{U}_i'\mathbf{V}$, and the observed outcomes under control, $\mathbf{Y}_i(\mathbf{0})^{obs}$, the treatment assignment is independent of the potential outcomes under control. This assumption is analogous to, but weaker than, the classic ignorability assumption required for causal inference, because it allows for unmeasured confounding to be captured either by observed variables, or by the time-varying latent variable. Formally, this assumption is written as 
\[
T_{0i} \independent \mathbf{Y}_i(\mathbf{0}) \; | \; \mathbf{Y}_i(\mathbf{0})^{obs}, \mathbf{Z}_i,\mathbf{U}_i^T\mathbf{V}
\]

\vspace{.5cm}

\textit{Assumption 3: Approximation of unobservables.} This assumption states that the matrix of unobserved features impacting the outcomes over time can be approximated through a low-rank matrix factorization. Equivalently, this requires that the unobserved time-varying confounders can be captured by a small number of latent factors, $K$, defined by the matrix factorization. This assumption is analogous to the assumption of correct model specification in standard regression models, but adapted to the context of latent variable models. Formally, this assumption requires that, for each unit, there is a latent (possibly time-varying) confounder of the treatment effect, $\mathbf{W}_i$, that can be approximated by a low-rank matrix factorization $\mathbf{W}_i=\mathbf{U}_i^T\mathbf{V}$.

\vspace{.5cm}

\textit{Assumption 4: Conditional exchangeability.} Conditional on any observed covariates $\mathbf{Z}$, and latent variable $\mathbf{W}$, elements of $\mathbf{Y(0)}$ are exchangeable. This assumption is needed in order to define the posterior predictive distribution used to estimate the missing values in $\mathbf{Y(0)}$.

\vspace{.5cm}

Under these assumptions,  \citeSM{pang2020bayesian} show that the causal effects are identified by Bayesian matrix completion models.

\clearpage

\subsection{Bayesian modularization}\label{s:s:modular}

Our modularized model can also be viewed as a complex Bayesian multiple imputation procedure. Causal inference in the potential outcomes framework is often treated as a missing data problem, i.e., the counterfactual outcomes are considered missing data that needs to be imputed. In our case, the TC-specific causal inference models are used to impute counterfactual outcomes, and the resulting treatment effects from all the TCs are passed into the predictive model. Although the causal and predictive models suggest incompatible distributions for the missing counterfactual outcomes, the use of incompatible imputation and analysis models has been well-studied. This approach, often referred to as a fully conditional model specification, incompatible Markov Chain Monte Carlo (MCMC), or multiple imputation with incongenial sources of input, has been shown to often perform better than fully Bayesian data augmentation with complex models \citepSM{meng1994multiple,rubin2003nested,schafer2003multiple,van2006fully,van2007multiple,kuo2018simulating}.

We introduce our modularization approach using this missing data perspective. Our goal is to illustrate the principles used to modularize the model by providing general forms of the modularized full conditional distributions of all parameters. For a given TC $s$, denote by $Y_{s}(0)$ the full set of potential outcomes under control, composed of both the observed values $Y_{s}^{obs}(0)$ and the missing values $Y_{s}^{mis}(0)$. The causal component of the model aims to estimate the $Y_{s}^{mis} (0)$. Denote the TC-specific causal parameters by $\phi_{s}$ and denote the treatment indicators for each unit/time collectively as $D_s$ (note that from the missing data perspective, the treatment indicators can be viewed as missingness indicators). We assume throughout this section that $Y_{s}(1)$, the (observed) potential outcomes under treatment for the treated units at post-treatment times, are fixed and known quantities. Then the IEEs, $\theta_s$, are assumed to be a deterministic function of the $Y_{s}^{mis} (0)$. In the predictive component of the model, let $X$ denote the predictors and $\mathbf{\beta}$ denote the predictive model parameters. (We again focus on a single outcome-specific model.)

Then the posterior distribution of the parameters and missing data can be expressed as
{\footnotesize
\[
P\left(\left\lbrace Y_s^{mis} (0)\right\rbrace_{s=1}^S, \left\lbrace\phi_s \right\rbrace_{s=1}^S, \mathbf{\beta} \; \; | \; \; \left\lbrace Y_s^{obs} (0) \right\rbrace_{s=1}^S, \left\lbrace  D_s\right\rbrace_{s=1}^S, X\right)
\]
}
Traditional Bayesian MCMC sampling algorithms successively sample from the full conditional posterior distribution of each of the unknown parameters. For the missing data, we obtain the following full conditional distribution
{\footnotesize
\begin{align*}
   P\left(\left\lbrace Y_s^{mis} (0)\right\rbrace_{s=1}^S
   \; \; | \; \; 
   \left\lbrace Y_s^{obs} (0) \right\rbrace_{s=1}^S,
   \left\lbrace  D_s\right\rbrace_{s=1}^S,
   X,
   \left\lbrace\phi_s \right\rbrace_{s=1}^S,
   \mathbf{\beta}\right)
    \propto &
   P\left(\left\lbrace Y_s^{obs} (0) \right\rbrace_{s=1}^S,
   \left\lbrace Y_s^{mis} (0) \right\rbrace_{s=1}^S,
   \left\lbrace  D_s\right\rbrace_{s=1}^S,
   X
   \; \; | \; \;
   \left\lbrace\phi_s \right\rbrace_{s=1}^S,
   \mathbf{\beta}\right) \\
    = & P\left(\left\lbrace D_s\right\rbrace_{s=1}^S
   \; \; | \; \;
   \left\lbrace Y_s^{obs} (0) \right\rbrace_{s=1}^S,
   \left\lbrace Y_s^{mis} (0) \right\rbrace_{s=1}^S,
   X,
   \left\lbrace\phi_s \right\rbrace_{s=1}^S,
   \mathbf{\beta} \right)\times\\
   & P\left( \left\lbrace Y_s^{obs} (0) \right\rbrace_{s=1}^S,
    \; \; | \; \;
     \left\lbrace Y_s^{mis} (0) \right\rbrace_{s=1}^S,
   X,
   \left\lbrace\phi_s \right\rbrace_{s=1}^S,
   \mathbf{\beta} \right) \times\\
   & P\left(\left\lbrace Y_s^{mis} (0) \right\rbrace_{s=1}^S
   \; \; | \; \;
   X,
   \left\lbrace\phi_s \right\rbrace_{s=1}^S,
   \mathbf{\beta} \right)\times \\
   & P\left( X
    \; \; | \; \;
   \left\lbrace\phi_s \right\rbrace_{s=1}^S,
   \mathbf{\beta} \right)
\end{align*}
}
Then if we assume that
{\footnotesize
\[\left\lbrace D_s, Y_s^{miss} (0), Y_s^{obs} (0) \right\rbrace \independent \left\lbrace D_{s'}, Y_{s'}^{miss} (0), Y_{s'}^{obs} (0) \right\rbrace_{s' \neq s} \; \; | \; \; \phi_s\]
}
we obtain
{\footnotesize
\begin{align*}
    \prod_{s=1}^S P\left( D_s \; \; | \; \;
    Y_s^{obs} (0), Y_s^{mis} (0),
   X,
   \phi_s ,
   \mathbf{\beta} \right)
   P\left( Y_s^{obs} (0) 
    \; \; | \; \;
    Y_s^{mis} (0) ,
   X,
   \phi_s ,
   \mathbf{\beta} \right)
    P\left( Y_s^{mis} (0) 
     \; \; | \; \;
   X,
   \phi_s ,
   \mathbf{\beta} \right)
    P\left( X
    \; \; | \; \;
   \left\lbrace\phi_s \right\rbrace_{s=1}^S,
   \mathbf{\beta} \right)
\end{align*}
}
Then assuming the causal identifying assumptions hold, we have unconfoundedness conditional on the causal model parameters, so that $P\left( D_s \; \; | \; \;
    Y_s^{obs} (0), Y_s^{mis} (0),
   X,
   \phi_s ,
   \mathbf{\beta} \right)=P\left( D_s \; \; | \; \;
   Y_s^{obs} (0),
   X,
   \phi_s ,
   \mathbf{\beta} \right)$, which does not depend on $Y_s^{mis} (0)$ and thus drops out. $P\left( X
    \; \; | \; \;
   \left\lbrace\phi_s \right\rbrace_{s=1}^S,
   \mathbf{\beta} \right)$ also does not depend on $Y_s^{mis} (0)$ and drops out. This leaves us with the following full conditional distribution
   {\footnotesize
   \begin{align*}
   P\left(\left\lbrace Y_s^{mis} (0)\right\rbrace_{s=1}^S
   \; \; | \; \; 
   \left\lbrace Y_s^{obs} (0) \right\rbrace_{s=1}^S,
   \left\lbrace  D_s\right\rbrace_{s=1}^S,
   X,
   \left\lbrace\phi_s \right\rbrace_{s=1}^S,
   \mathbf{\beta}\right)
    \propto 
        \prod_{s=1}^S 
   P\left( Y_s^{obs} (0) 
    \; \; | \; \;
    %Y_s^{mis} (0) ,
   X,
   \phi_s ,
   \mathbf{\beta} \right)
    P\left( Y_s^{mis} (0) 
     \; \; | \; \;
   X,
   \phi_s ,
   \mathbf{\beta} \right)
    \end{align*}
    }
This distribution factorizes so that the $ Y_s^{mis} (0)$ can be sampled separately for each TC. Moreover, conditional on the causal model parameters, $ Y_s^{obs} (0)$ and $ Y_s^{mis} (0)$ do not depend on $X$ or $\mathbf{\beta}$ so that 
{\footnotesize
  \begin{align*}
   P\left(\left\lbrace Y_s^{mis} (0)\right\rbrace_{s=1}^S
   \; \; | \; \; 
   \left\lbrace Y_s^{obs} (0) \right\rbrace_{s=1}^S,
   \left\lbrace  D_s\right\rbrace_{s=1}^S,
   X,
   \left\lbrace\phi_s \right\rbrace_{s=1}^S,
   \mathbf{\beta}\right)
    \propto 
        \prod_{s=1}^S 
   P\left( Y_s^{obs} (0) 
    \; \; | \; \;
    Y_s^{mis} (0) ,
   \phi_s  \right)
    P\left( Y_s^{mis} (0) 
     \; \; | \; \;
   \phi_s \right)
    \end{align*}
    }
and sampling of $Y_s^{mis}(0)$ can be conducted separately for each TC and without integrating feedback from the predictive component of the model. The derivation of the full conditional distribution for the $\left\lbrace\phi_s \right\rbrace_{s=1}^S$ proceeds similarly, i.e.,
{\footnotesize
\begin{align*}
   P\left(\left\lbrace\phi_s \right\rbrace_{s=1}^S,
   \; \; | \; \; 
   \left\lbrace Y_s^{obs} (0) \right\rbrace_{s=1}^S,
   \left\lbrace Y_s^{mis} (0)\right\rbrace_{s=1}^S,
   \left\lbrace  D_s\right\rbrace_{s=1}^S,
   X,
   \mathbf{\beta}\right)
    = &
   P\left(\left\lbrace Y_s^{obs} (0) \right\rbrace_{s=1}^S,
   \left\lbrace Y_s^{mis} (0) \right\rbrace_{s=1}^S,
   \left\lbrace  D_s\right\rbrace_{s=1}^S,
   X,
   \left\lbrace\phi_s \right\rbrace_{s=1}^S,
   \mathbf{\beta}\right) \\
    = &  P\left( \left\lbrace Y_s^{obs} (0) \right\rbrace_{s=1}^S, \left\lbrace Y_s^{mis} (0) \right\rbrace_{s=1}^S, \left\lbrace D_s\right\rbrace_{s=1}^S
    \; \; | \; \;
   X,
   \left\lbrace\phi_s \right\rbrace_{s=1}^S,
   \mathbf{\beta} \right) \times\\
   & P\left( X
    \; \; | \; \;
   \left\lbrace\phi_s \right\rbrace_{s=1}^S,
   \mathbf{\beta} \right) \times P\left( \mathbf{\beta} \; \; | \; \;
   \left\lbrace\phi_s \right\rbrace_{s=1}^S \right)\times P(\left\lbrace\phi_s \right\rbrace_{s=1}^S)\\
   \propto & P\left( X
    \; \; | \; \;
   \left\lbrace\phi_s \right\rbrace_{s=1}^S,
   \mathbf{\beta} \right) \times P\left( \mathbf{\beta} \; \; | \; \;
   \left\lbrace\phi_s \right\rbrace_{s=1}^S \right) \\
   & \prod_{s=1}^S P\left( Y_s^{obs} (0) , Y_s^{mis} (0) %, D_s
    \; \; | \; \;
   \phi_s ,
   \mathbf{\beta} \right) P(\phi_s)
\end{align*}
}
Following conventions in the modularization literature, the $P\left( X
    \; \; | \; \;
   \left\lbrace\phi_s \right\rbrace_{s=1}^S,
   \mathbf{\beta} \right)P\left( \mathbf{\beta} \; \; | \; \;
   \left\lbrace\phi_s \right\rbrace_{s=1}^S \right)$ term can be dropped from this full conditional to prevent feedback, and sampling of the $\phi_s$ can proceed separately by TC and disjoint from the predictive model.
   
   Finally, the full conditional for $\mathbf{\beta}$ can be expressed as
   {\footnotesize
   \begin{align*}
   P\left(\mathbf{\beta}
   \; \; | \; \; 
   \left\lbrace Y_s^{obs} (0) \right\rbrace_{s=1}^S,
   \left\lbrace Y_s^{mis} (0)\right\rbrace_{s=1}^S,
   \left\lbrace  D_s\right\rbrace_{s=1}^S,
   X,
   \left\lbrace\phi_s \right\rbrace_{s=1}^S\right)
    \propto &
   P\left(\left\lbrace Y_s^{obs} (0) \right\rbrace_{s=1}^S,
   \left\lbrace Y_s^{mis} (0) \right\rbrace_{s=1}^S,
   \left\lbrace  D_s\right\rbrace_{s=1}^S,
   X,
   \left\lbrace\phi_s \right\rbrace_{s=1}^S,
   \mathbf{\beta}\right) \\
    = &  P\left( \left\lbrace Y_s^{obs} (0) \right\rbrace_{s=1}^S, \left\lbrace Y_s^{mis} (0) \right\rbrace_{s=1}^S, \left\lbrace D_s\right\rbrace_{s=1}^S
    \; \; | \; \;
   X,
   \left\lbrace\phi_s \right\rbrace_{s=1}^S,
   \mathbf{\beta} \right) \times\\
   & P\left( X
    \; \; | \; \;
   \left\lbrace\phi_s \right\rbrace_{s=1}^S,
   \mathbf{\beta} \right)\times P\left( \mathbf{\beta} \; \; | \; \;
   \left\lbrace\phi_s \right\rbrace_{s=1}^S \right) \times P(\left\lbrace\phi_s \right\rbrace_{s=1}^S)\\
   \propto & P\left( X
    \; \; | \; \;
   \left\lbrace\phi_s \right\rbrace_{s=1}^S,
   \mathbf{\beta} \right) P\left( \mathbf{\beta} \; \; | \; \;
   \left\lbrace\phi_s \right\rbrace_{s=1}^S \right)
\end{align*}
}
Thus $\mathbf{\beta}$ is sampled conditional on the causal model parameters, allowing information from the causal models to flow into the predictive model. This is the rationale that guides our modularization approach.

\begin{figure}[h!]
\centering
\includegraphics[scale=.5]{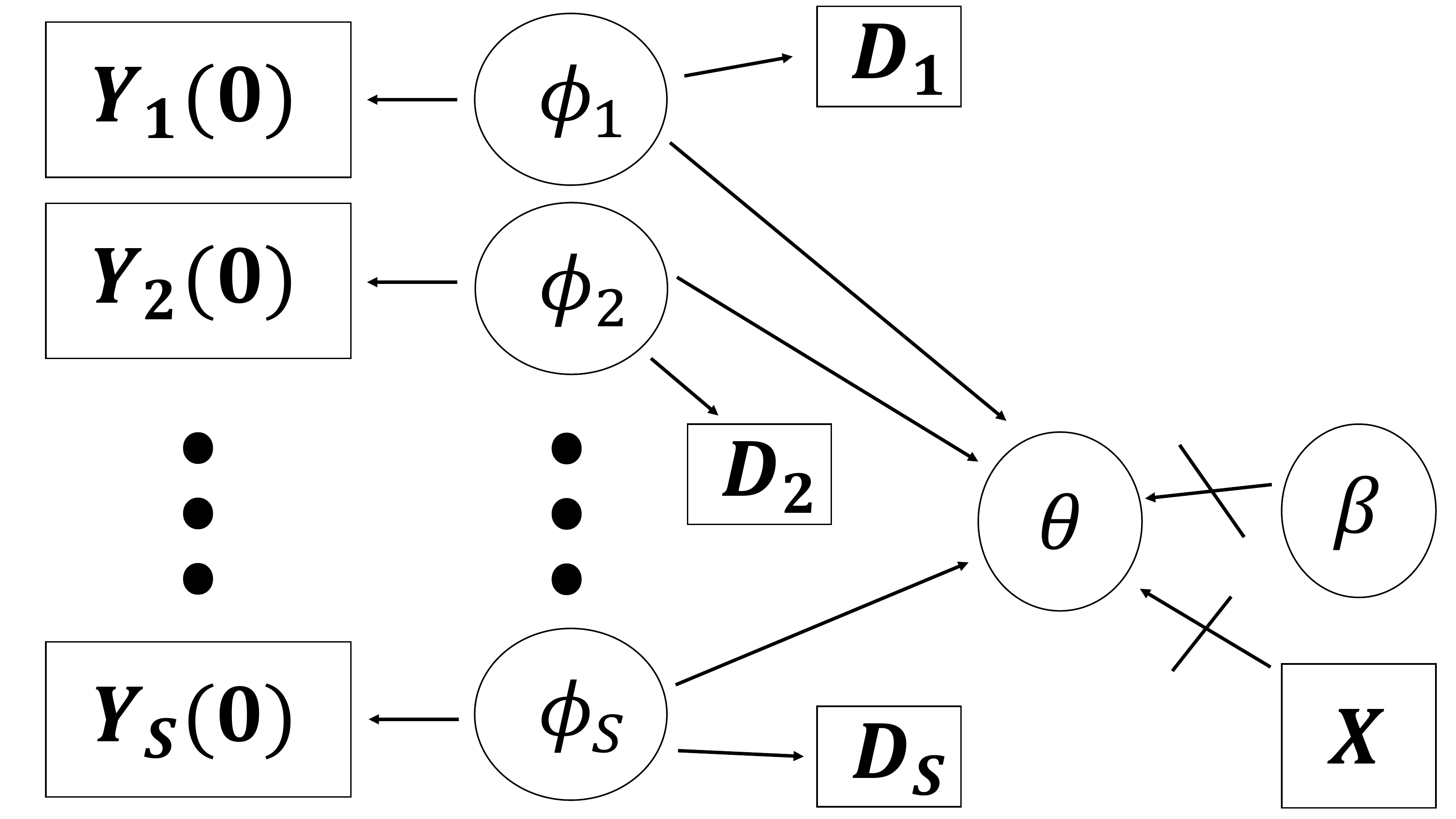}
\caption{DAG for Bayesian modularization}
\label{fig:bayesmod}
\end{figure}

Our full modularized model can be represented as the directed graphical model in Figure~\ref{fig:bayesmod}. The one-way flow of information from the causal sub-models to the predictive model is depicted by the cuts in the arrows in Figure~\ref{fig:bayesmod}. The MCMC sampling algorithm proceeds as follows. First, for each TC, $s \in \lbrace 1, \dots,  S \rbrace$, we fit our causal sub-model, drawing a posterior sample of the IEE for each affected county denoted by $\theta_{is}^{(m)}$ for county $i \in W_{s}$ and posterior sample $m$ ($m=1,...,M)$. We then update the predictive model parameters conditional on the IEE draw. We draw a posterior sample, $\mathbf{\beta}^{(m)}$, of the parameters from this predictive model and iterate this procedure until convergence of all parameters in all models.% The posterior samples from the predictive model can then be used to make county-specific predictions of the number of health events that will be caused by an impending TC. 

\clearpage

\subsection{Population displacement}\label{s:s:pop_displace}
The zipcode/county of residence recorded in Medicare claims data for each recipient is updated yearly and represents their place of residence at either the end of the specified year or early the following year (timing changes slightly year-to-year). This implies that, if a person moves from county A to county B during a given year, all their hospitalizations for that year will be assigned to county B. Because the Atlantic hurricane season occurs primarily in the latter half of the year (June-November), use of the county of residence at year's end for exposure classification and population denominator construction is generally appropriate.

Yet, for a few severe storms that resulted in substantial long-term population displacement, such as Hurricane Katrina \citepSM{deryugina2018does}, some degree of exposure misclassification is likely. However, we would expect that the impact of any such exposure misclassification would be to bias IEEs towards the null. Consider a scenario in which county A is exposed to a TC and county B is a control for that TC. Then, if a person living in county A at the time of exposure is hospitalized due to the TC, but is displaced to county B before the end of the year, their TC-attributable hospitalizations will be assigned to control county B. The resulting increase in adverse event rates in control counties should lead to an upward bias in the estimated $Y_{it}(0), \; t\geq T_0$ and a corresponding downward bias in the IEE for the exposed counties. In spite of this potential for underestimated IEEs, we estimate large adverse health impacts for Hurricane Katrina. For most other TCs assessed by our study, we would expect that TC-related long-term displacement rates are low, based on previous work that found that, even for unusually strong TCs (Category 3 or greater at landfall), outward migration increases on average by only 6.85\% in the year following the storms \citepSM{ouattara2014hurricane}.

\clearpage

\subsection{Selection of $K$}\label{s:s:select_k}

\begin{figure}[h!]
\centering
\includegraphics[scale=.5]{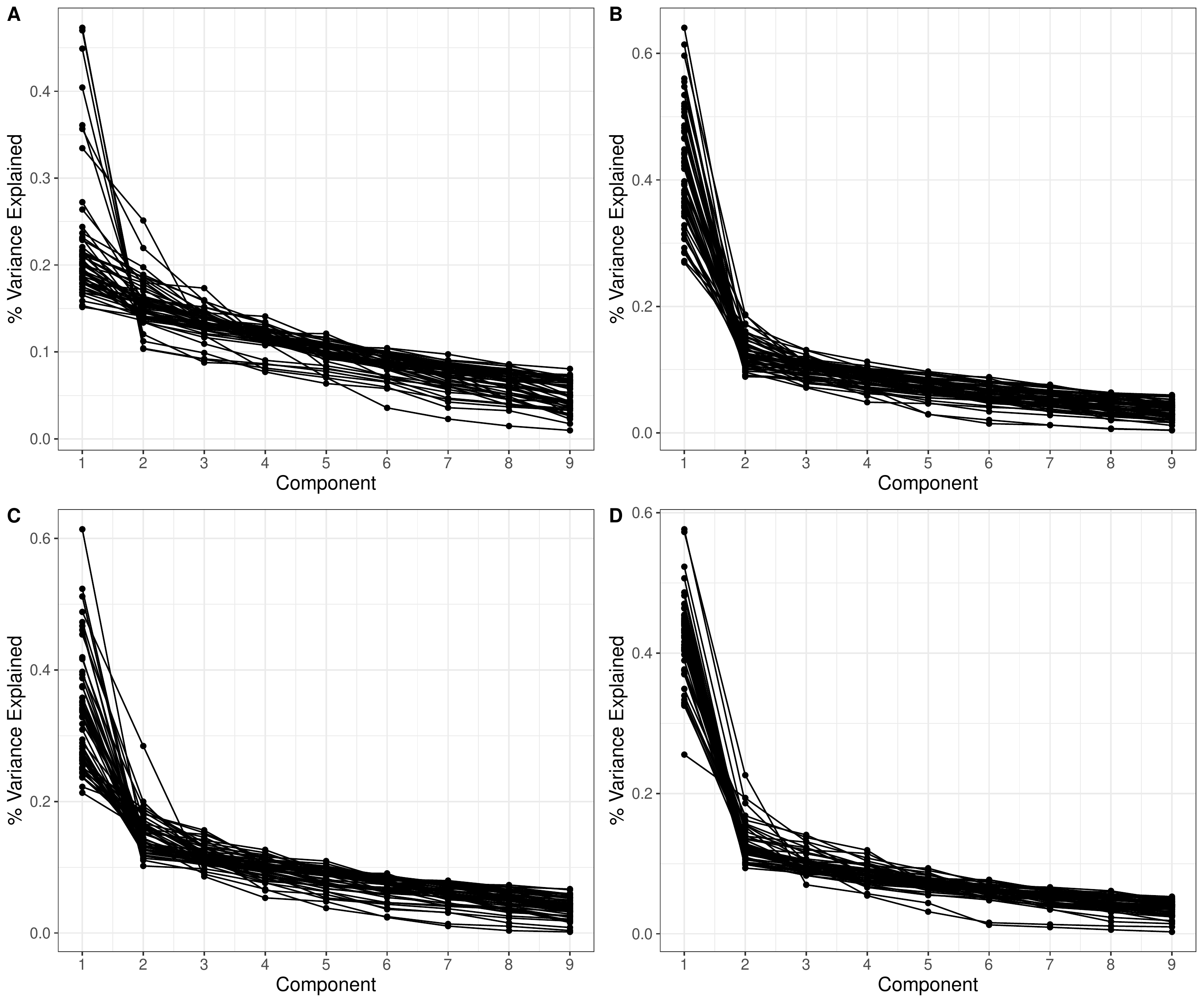}
\caption{Scree plots for PCA fit to the panel data matrix for each TC for mortality (A), respiratory hospitalizations (B), COPD hospitalizations (C), and CVD hospitalizations (D).}
\label{s:fig:select_k}
\end{figure}

We wish to select a $K$ value that preserves approximately 70\% of the variance in the panel data matrix for all TCs and all outcomes. 70\% is chosen because it represents a compromise between retaining variability necessary for accurate predictions, while not overfitting to the data. To explore the choice of $K$ that best satisfies this criterion, we conduct exploratory principal component analyses (PCA) on the panel data matrices. Specifically, we remove the final column of the panel data matrix for each TC and outcome (because it contains missing values which are not permitted in PCA), and we implement PCA on each resulting matrix. The scree plots showing the percent of variance explained by each prinicipal component (for each panel data matrix) are shown in Figure~\ref{s:fig:select_k}. On average across TCs, the first four principal components explain 63\% of the variance in mortality, 73\% in respiratory hospitalizations, 70\% in COPD hospitalizations, and 74\% in CVD hospitalizations. Thus $K=4$ appears to be a reasonable choice to preserve around 70\% of the variability in the panel data matrices. Moreover, as demonstrated by Figure~\ref{s:fig:select_k}, each principal component beyond the fourth generally explains less than 10\% of the variance in the panel data.

\clearpage

\subsection{Predictive model selection}\label{s:s:pred_mod_select}
Candidate predictive models are Bayesian additive regression trees \citepSM{chipman2010bart} (BART) and numerous variants of a linear regression model. In the regression models, we always include a restricted cubic spline on year to allow for flexible time trends in health effects of TCs. We consider regression models with (1) all additional predictors included as linear terms, (2) a restricted cubic spline on TC maximum sustained windspeed and all additional predictors included as linear terms, and (3) interactions between all predictors and a binary indicator of hurricane-speed winds (sustained windspeeds $>33$m/s). We test each predictive model candidate both with and without state included as a predictor.

To evaluate and compare predictive models, we fit them by plugging in the county-level excess rate point estimates as the outcomes (without passing their full posterior distributions into the predictive model). Using five-fold cross-validation, we compare out-of-sample predictive performance of these models. Root mean square error for each model is given in Table~\ref{s:tab:cv_results}. 
For most outcomes, we find similar predictive performance across models. Overall, we find the most consistently strong performance for the regression models with splines on windspeed, excluding state.
Thus, for the primary predictive sub-model, we select the Bayesian linear model with a spline on windspeed, excluding state as a predictor. For interpretability, we also provide results from a linear regression model without the windspeed spline.

% latex table generated in R 3.6.3 by xtable 1.8-4 package
% Wed Sep  9 19:10:15 2020
\begin{table}[ht]
\centering
\caption{Root mean square error from 5-fold cross-validation for each considered predictive model.}
\label{s:tab:cv_results}
\begin{tabular}{rrrrr}
  \hline
 & Mortality & Resp & COPD & CVD \\ 
  \hline
Linear with state & 90.15 & 64.24 & 41.82 & 96.99 \\ 
  Linear without state & 89.96 & 64.34 & 42.08 & 95.84 \\ 
  Spline with state & 89.92 & \textbf{64.15} & 41.88 & 96.99 \\ 
  Spline without state & \textbf{89.59} & 64.25 & 42.14 & \textbf{95.83} \\ 
  Linear, hurricane stratified & 92.15 & 64.57 & 42.41 & 96.04 \\ 
  Spline, hurricane stratified & 92.17 & 64.57 & 42.45 & 96.05 \\ 
  BART with state & 90.85 & 64.28 & \textbf{41.53} & 95.85 \\ 
  BART without state & 101.36 & 64.26 & 41.97 & 96.07 \\ 
   \hline
\end{tabular}
\end{table}

\clearpage

\subsection{Sensitivity analyses}\label{s:s:sensitivity}

\subsubsection{Causal models}
\begin{figure}[h!]
\centering
\includegraphics[scale=.75]{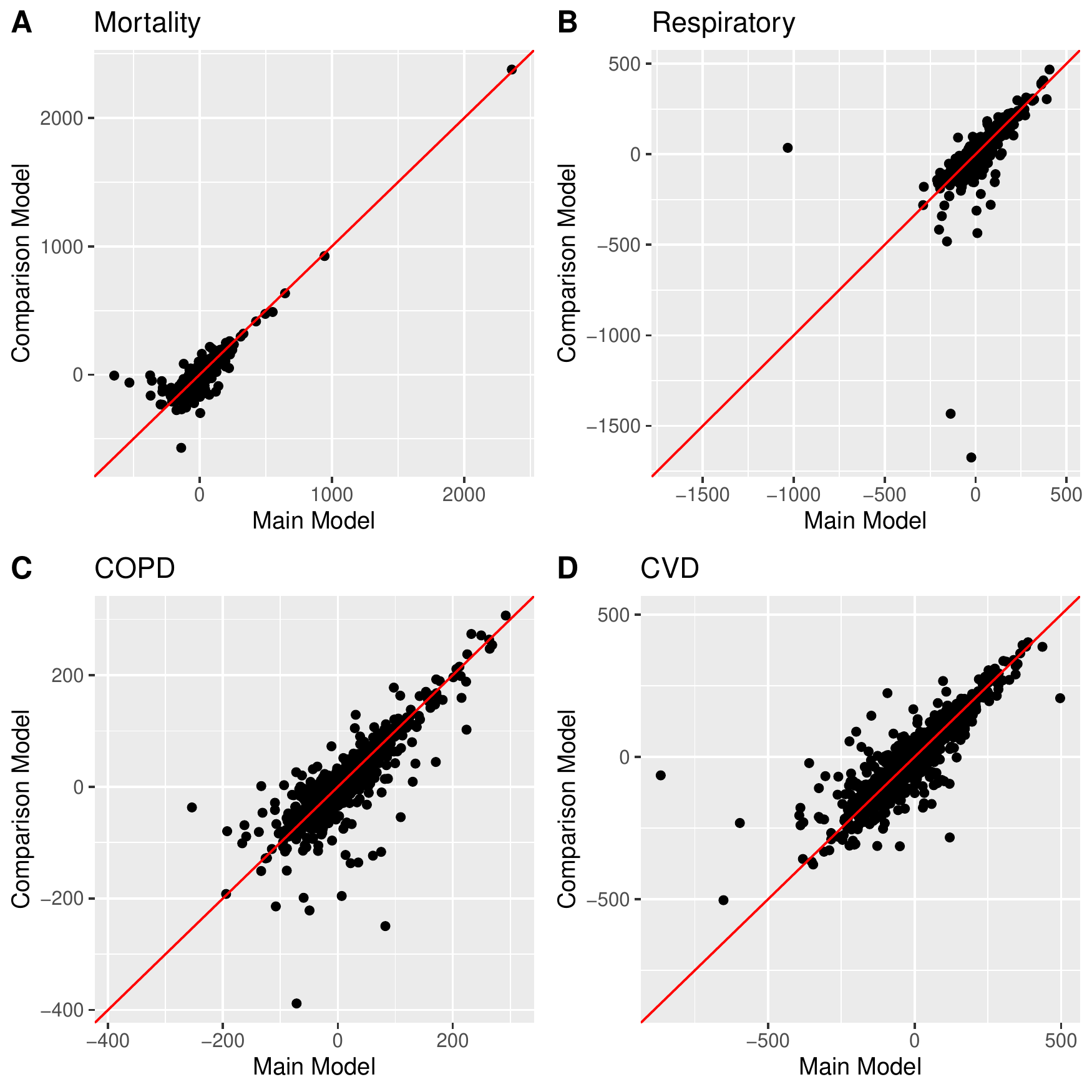}
\caption{County-level excess rate estimates from the main causal models compared to causal models with counties adjacent to TC-exposed counties removed.}
\label{s:fig:comp_rmadj}
\end{figure}

Because TC exposures are complex, our use of a binary TC exposure metric in the causal models could lead to some degree of exposure misclassification. In our models, we use as ``controls'' counties that are classified as unexposed and are within 150 miles of at least one exposed county. Because these controls counties lie near the path of the TC, it is likely that some of them received impacts. We anticipate that, if anything, including these counties as controls would lead to conservative results (i.e., treatment effects that are closer to the null). However, to evaluate the degree of influence this misclassification could have on our results, we fit causal models excluding from each TC's model any control county that is adjacent to an exposed county. We expect that the counties nearest to exposed counties would be most likely to experience TC impacts. Figure~\ref{s:fig:comp_rmadj} compares the resulting county-level excess rate estimates with the estimates from our primary models. The estimates from the two models are highly clustered around the one-to-one line for all outcomes, with no clear systematic differences. This suggests that the causal models are robust to TC exposure misclassification.

\begin{figure}[h!]
\centering
\includegraphics[scale=.75]{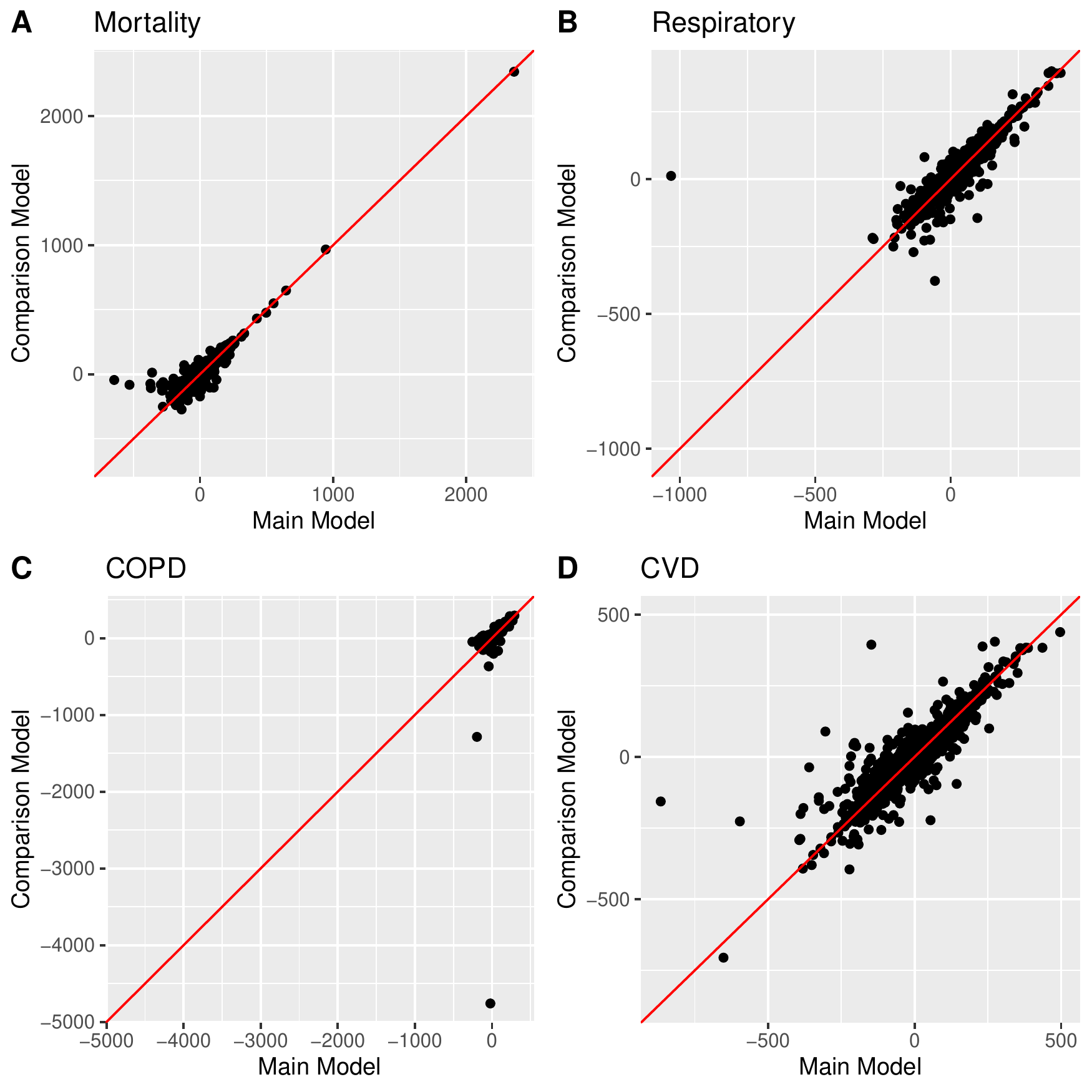}
\caption{County-level excess rate estimates from the main causal models ($K=4$) compared to causal models with $K=3$.}
\label{s:fig:comp_k3}
\end{figure}

\begin{figure}[h!]
\centering
\includegraphics[scale=.75]{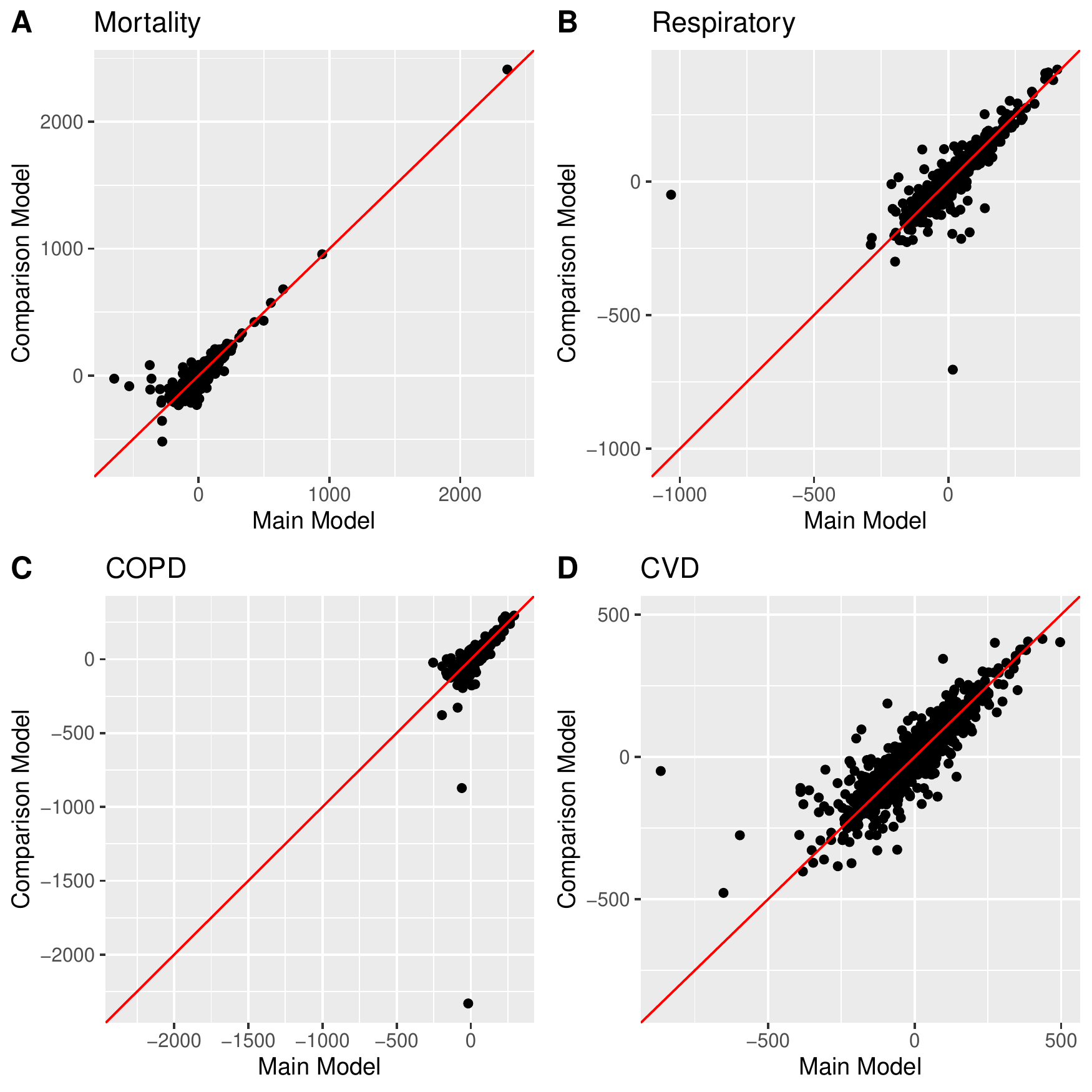}
\caption{County-level excess rate estimates from the main causal models ($K=4$) compared to causal models with $K=5$.}
\label{s:fig:comp_k5}
\end{figure}

Moreover, in the causal models, results could be sensitive to the specification of the number of latent factors, $K$. Thus, we compare the county-level excess rate estimates obtained when specifying $K=\left\lbrace 3,5 \right\rbrace$ to the estimates from our main model with $K=4$. The comparison of the estimates from these models is shown in Figures~\ref{s:fig:comp_k3} and~\ref{s:fig:comp_k5}. As in the previous sensitivity analysis, estimates from the models are highly clustered around the one-to-one line for all outcomes, with no clear systematic differences. This suggests that our results are robust to the specification of $K$.

\clearpage

\subsubsection{Predictive model}
\begin{figure}[h!]
\centering
\includegraphics[scale=.75]{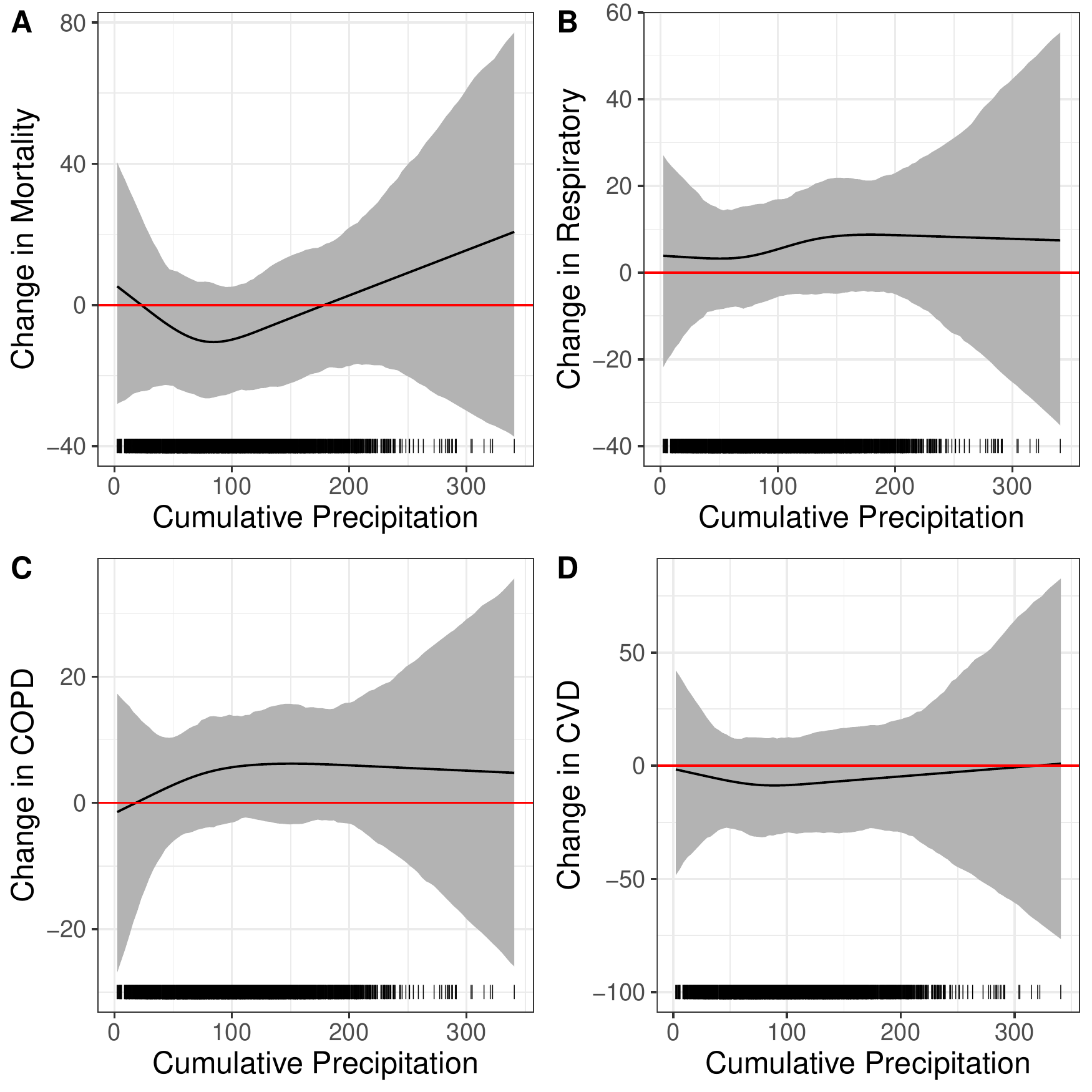}
\caption{Relationship between cumulative precipitation and excess rate per 100,000 of mortality (A), respiratory hospitalizations (B), COPD hospitalizations (C), and CVD hospitalizations (D).}
\label{s:fig:cumulative_precip}
\end{figure}

TCs can bring extreme rainfall amounts and flooding, and we hypothesize that these exposures may cause increases in some adverse health events. Currently, the \texttt{hurricaneexposuredata} package contains TC rainfall data only through the year 2011. Due to the missing precipitation data for later years, we do not include this variable in our main predictive models, but here, as an additional sensitivity analysis, we fit the predictive models only using TCs in years 2011 and prior with precipitation as a predictor. As in the primary models, we fit regression models excluding state indicators but including all other predictors, with restricted cubic splines on windspeed and year. We also include a restricted cubic spline on county-level cumulative TC precipitation, measured over a four-day window starting two days prior to the storm's closest approach to the county. The precipitation splines from the fully modularized models are shown in Figure~\ref{s:fig:cumulative_precip}. After adjusting for all the other predictors, there appears to be weak or no effect of precipitation on any of the health outcomes evaluated here. Thus, the inference and predictions from our model are unlikely to be sensitive to the inclusion of precipitation information.

\clearpage

\subsection{Additional tables and figures}\label{s:s:additional}
\hspace{5cm}

\begin{table}[ht]
\centering
\caption{Definition of estimands. $s=\left\lbrace1,...,S\right\rbrace$ indexes TCs and $i\in W_s$ indexes treated counties for TC $s$.}
\centerline{
\begin{tabular}{lp{6cm}p{6cm}}
  \hline
Name & Definition & Formula \\ 
  \hline
  Individual excess events (IEE) & County-level excess events attributable to a single TC $s$ over the treatment period & $\theta_{si}=\sum_{t\geq T_0} \left[ Y_{sit} (1)-Y_{sit} (0) \right]$\\
  Individual excess rate & County-level excess event rate (per 100,000 population) attributable to a single TC $s$ over the treatment period & $\theta^*_{si}=100000\times (\theta_{si}/p_{iT})$\\
  TC-specific excess events & Cumulative excess events summed across all counties impacted by a single TC $s$ & $\sum_{i\in W_s} \theta_{si}$\\
  TC-specific excess rate & Excess event rate (per 100,000 population) across all counties impacted by a single TC $s$ & $100000\times (\sum_{i\in W_s} p_{iT})^{-1} \sum_{i\in W_s} \theta_{si}$\\
  Total excess events (TEE) & Cumulative TC-attributable excess events summed over all TCs and counties in the study & $TEE= \sum_{s=1}^S \sum_{i\in W_s} \theta_{si}$\\
  Average excess rate (AER) & Average of the TC-attributable excess rates across all county-level TC exposures in the study & $AER=\frac{1}{N_{total}}\sum_{s=1}^S \sum_{i \in W_s} \theta_{si}^*$, $N_{total} = \sum_{s=1}^S |W_s|$ is the total number of county-level TC exposures in our analyses\\
   \hline
\end{tabular}}
\label{s:tab:estimands}
\end{table}

% latex table generated in R 3.5.2 by xtable 1.8-4 package
% Fri Sep  4 10:40:30 2020
\begin{table}[h!]
\centering
\caption{Name and year of each TC included in the study, the number of treated and control counties for the TC, and the rates (per 100,000) of each health outcome in Medicare in the treated and control counties during the 140 day period surrounding the TC.}
\label{s:tab:storm_feat}
\footnotesize
\centerline{
\begin{tabular}{lrrrrrrrrrr}
  \hline
TC & N Trt & N Ctl & Mort Trt & Mort Ctl & Resp Trt & Resp Ctl & COPD Trt & COPD Ctl & CVD Trt & CVD Ctl \\ 
  \hline
Alberto-2006 &  21 & 130 & 2036 & 1901 & 1361 & 1088 & 494 & 413 & 3210 & 2779 \\ 
  Alex-2004 &  18 &  86 & 1828 & 1858 & 1052 & 909 & 380 & 331 & 2882 & 2868 \\ 
  Allison-2001 &  22 & 143 & 2161 & 2209 & 1483 & 2050 & 584 & 743 & 3433 & 3832 \\ 
  Ana-2015 &   3 &  52 & 1476 & 1751 & 800 & 875 & 369 & 355 & 1838 & 1971 \\ 
  Andrea-2013 &  80 & 223 & 1718 & 1667 & 899 & 850 & 373 & 356 & 2059 & 1897 \\ 
  Arlene-2005 &   4 &  61 & 1871 & 2189 & 1472 & 1805 & 561 & 669 & 3010 & 3288 \\ 
  Arthur-2014 &  45 & 113 & 1642 & 1541 & 618 & 607 & 248 & 251 & 1864 & 1579 \\ 
  Barry-2001 &  18 &  59 & 1971 & 2099 & 1412 & 1397 & 556 & 554 & 3418 & 3552 \\ 
  Barry-2007 &   8 &  73 & 1778 & 1841 & 729 & 947 & 272 & 370 & 2187 & 2616 \\ 
  Beryl-2012 &  22 & 114 & 1638 & 1664 & 853 & 897 & 366 & 409 & 1996 & 2118 \\ 
  Bill-2003 &  16 &  77 & 2139 & 2056 & 1400 & 1581 & 546 & 554 & 3786 & 3495 \\ 
  Bill-2015 &   9 &  47 & 1793 & 1585 & 929 & 852 & 343 & 322 & 1877 & 1708 \\ 
  Bret-1999 &  15 &  28 & 1611 & 1903 & 1015 & 995 & 386 & 362 & 3101 & 2863 \\ 
  Charley-2004 &  76 & 120 & 1758 & 1801 & 849 & 918 & 341 & 333 & 2980 & 2804 \\ 
  Cindy-2005 &  13 &  78 & 2330 & 2004 & 1357 & 1562 & 489 & 560 & 3247 & 3086 \\ 
  Claudette-2003 &  36 &  41 & 1914 & 1866 & 1163 & 1433 & 399 & 531 & 3018 & 3370 \\ 
  Claudette-2009 &   3 &  53 & 1700 & 1787 & 1056 & 996 & 489 & 464 & 2713 & 2535 \\ 
  Dennis-1999 &  23 &  94 & 1792 & 1864 & 988 & 904 & 427 & 356 & 3058 & 2859 \\ 
  Dennis-2005 &  35 &  81 & 1990 & 2105 & 1452 & 1360 & 541 & 506 & 3254 & 2858 \\ 
  Dolly-2008 &  13 &  22 & 1617 & 1659 & 1123 & 783 & 429 & 290 & 2536 & 2088 \\ 
  Edouard-2008 &  10 &  72 & 2025 & 1767 & 1216 & 966 & 502 & 400 & 2565 & 2435 \\ 
  Ernesto-2006 &  72 & 192 & 1698 & 1734 & 787 & 789 & 307 & 296 & 2494 & 2653 \\ 
  Fay-2002 &   2 &  47 & 2047 & 1853 & 796 & 995 & 189 & 381 & 3541 & 3010 \\ 
  Fay-2008 &  52 &  50 & 1634 & 1794 & 763 & 884 & 354 & 409 & 2321 & 2370 \\ 
  Floyd-1999 & 142 & 156 & 1765 & 1824 & 904 & 890 & 315 & 325 & 2776 & 2913 \\ 
  Frances-2004 &  50 &  39 & 1734 & 1843 & 842 & 974 & 352 & 391 & 2960 & 3178 \\ 
  Gabrielle-2001 &  23 &  26 & 1802 & 1827 & 851 & 930 & 349 & 388 & 3073 & 3256 \\ 
  Gabrielle-2007 &   6 &  71 & 1696 & 1770 & 763 & 715 & 360 & 273 & 2292 & 2413 \\ 
  Gaston-2004 &  25 & 116 & 1761 & 1788 & 879 & 819 & 336 & 281 & 2855 & 2701 \\ 
  Gordon-2000 &  12 &  63 & 1862 & 1775 & 817 & 826 & 371 & 343 & 3018 & 3111 \\ 
  Gustav-2008 &  48 &  77 & 1844 & 1761 & 1062 & 914 & 431 & 392 & 2811 & 2341 \\ 
  Hanna-2002 &   2 &  74 & 1871 & 1973 & 960 & 1198 & 388 & 478 & 3397 & 3339 \\ 
  Hanna-2008 & 127 & 163 & 1660 & 1681 & 763 & 759 & 325 & 321 & 2470 & 2202 \\ 
  Harvey-1999 &   3 &  20 & 1876 & 1739 & 1309 & 823 & 535 & 334 & 3300 & 2950 \\ 
  Helene-2000 &  14 &  96 & 1988 & 1913 & 1089 & 960 & 418 & 381 & 3139 & 2983 \\ 
  Hermine-2010 &  13 &  30 & 1515 & 1599 & 891 & 742 & 369 & 295 & 2063 & 1897 \\ 
  Humberto-2007 &  31 &  81 & 1766 & 1838 & 942 & 956 & 334 & 329 & 2490 & 2580 \\ 
  Ike-2008 & 215 &  64 & 1782 & 1756 & 924 & 874 & 411 & 350 & 2471 & 2414 \\ 
  Irene-1999 &  28 & 125 & 1777 & 1820 & 1015 & 875 & 411 & 360 & 3119 & 2889 \\ 
  Irene-2011 & 146 & 146 & 1622 & 1623 & 793 & 755 & 329 & 313 & 2046 & 1884 \\ 
  Isaac-2012 &  41 &  70 & 1769 & 1785 & 888 & 936 & 343 & 393 & 2068 & 2062 \\ 
  Isabel-2003 & 135 &  86 & 1857 & 1890 & 1008 & 1036 & 362 & 378 & 2778 & 3079 \\ 
  Isidore-2002 &  25 &  80 & 2025 & 1981 & 1200 & 1221 & 486 & 463 & 3627 & 3291 \\ 
  Ivan-2004 &  48 &  72 & 1867 & 1959 & 1108 & 989 & 472 & 376 & 3428 & 3018 \\ 
  Jeanne-2004 &  45 &  35 & 1731 & 1830 & 834 & 874 & 352 & 346 & 2977 & 3020 \\ 
  Katrina-2005 &  78 &  72 & 1875 & 1781 & 957 & 902 & 370 & 338 & 2726 & 2856 \\ 
  Lee-2011 &  30 &  91 & 1769 & 1785 & 841 & 969 & 372 & 406 & 2159 & 2136 \\ 
  Lili-2002 &  34 &  86 & 1982 & 1962 & 1248 & 1148 & 452 & 447 & 3565 & 3228 \\ 
  Ophelia-2005 &  23 & 112 & 1687 & 1728 & 872 & 819 & 344 & 307 & 2782 & 2629 \\ 
  Rita-2005 &  40 &  89 & 1803 & 1823 & 989 & 866 & 381 & 322 & 2760 & 2687 \\ 
  Sandy-2012 & 105 & 121 & 1606 & 1671 & 683 & 718 & 288 & 297 & 1911 & 1843 \\ 
  Tammy-2005 &  12 &  52 & 1789 & 1734 & 873 & 793 & 313 & 315 & 2847 & 2697 \\ 
  Wilma-2005 &  18 &  21 & 1689 & 1735 & 770 & 767 & 305 & 310 & 2603 & 2714 \\ 
   \hline
\end{tabular}}
\end{table}

\begin{figure}[h!]
\centering
\includegraphics[scale=.75]{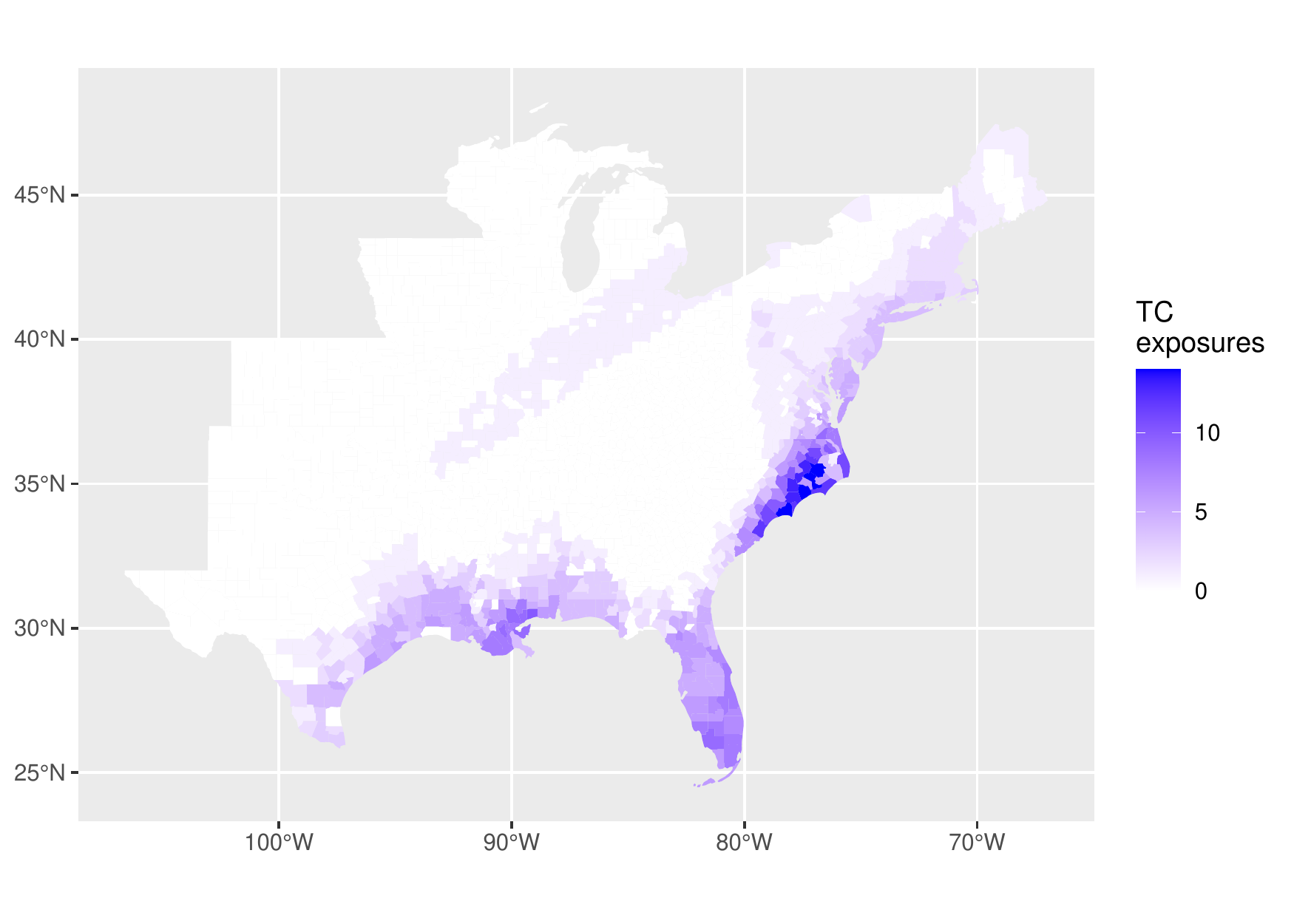}
\caption{Number of Atlantic Basin TC exposures included in our analyses (by county), 1999-2015.}
\label{s:fig:tc_hits}
\end{figure}

\begin{figure}[h!]
\centering
\centerline{\includegraphics[scale=.65]{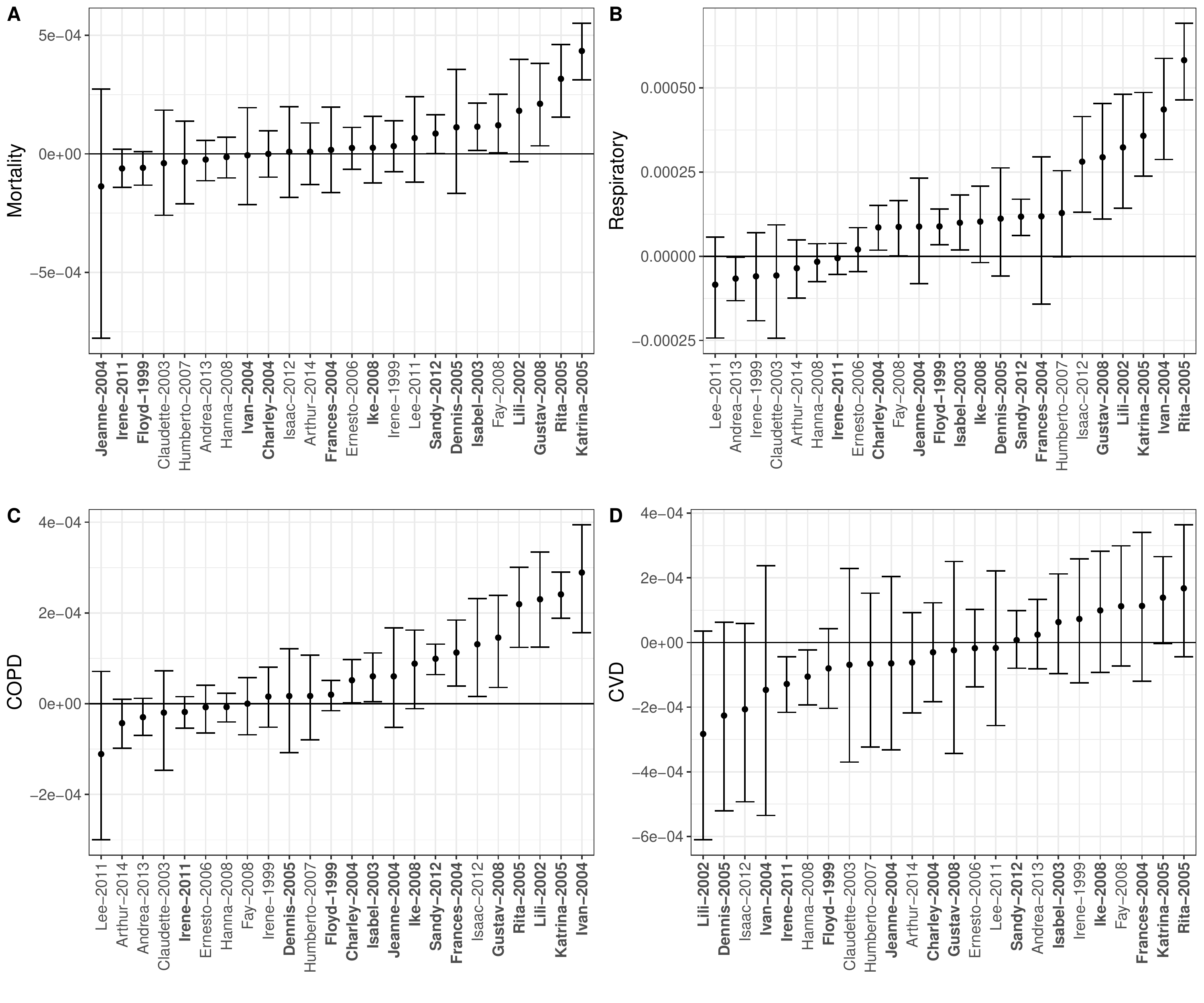}}
\caption{TC-specific excess rate estimates and 95\% predictive intervals for mortality (A), respiratory hospitalizations (B), COPD hospitalizations (C), and CVD hospitalizations (D) for TCs that impacted $>25$ counties. The TC-specific excess rate is the rate of excess events across the total population impacted by the TC. Bolded TC labels indicate storm names that were subsequently retired-- retirement occurs when a TC is so destructive that re-using the name is considered to be insensitive \protect\citepSM{nhc2020}.}
\label{s:fig:att}
\end{figure}

\begin{figure}[h!]
\centering
\centerline{\includegraphics[scale=.65]{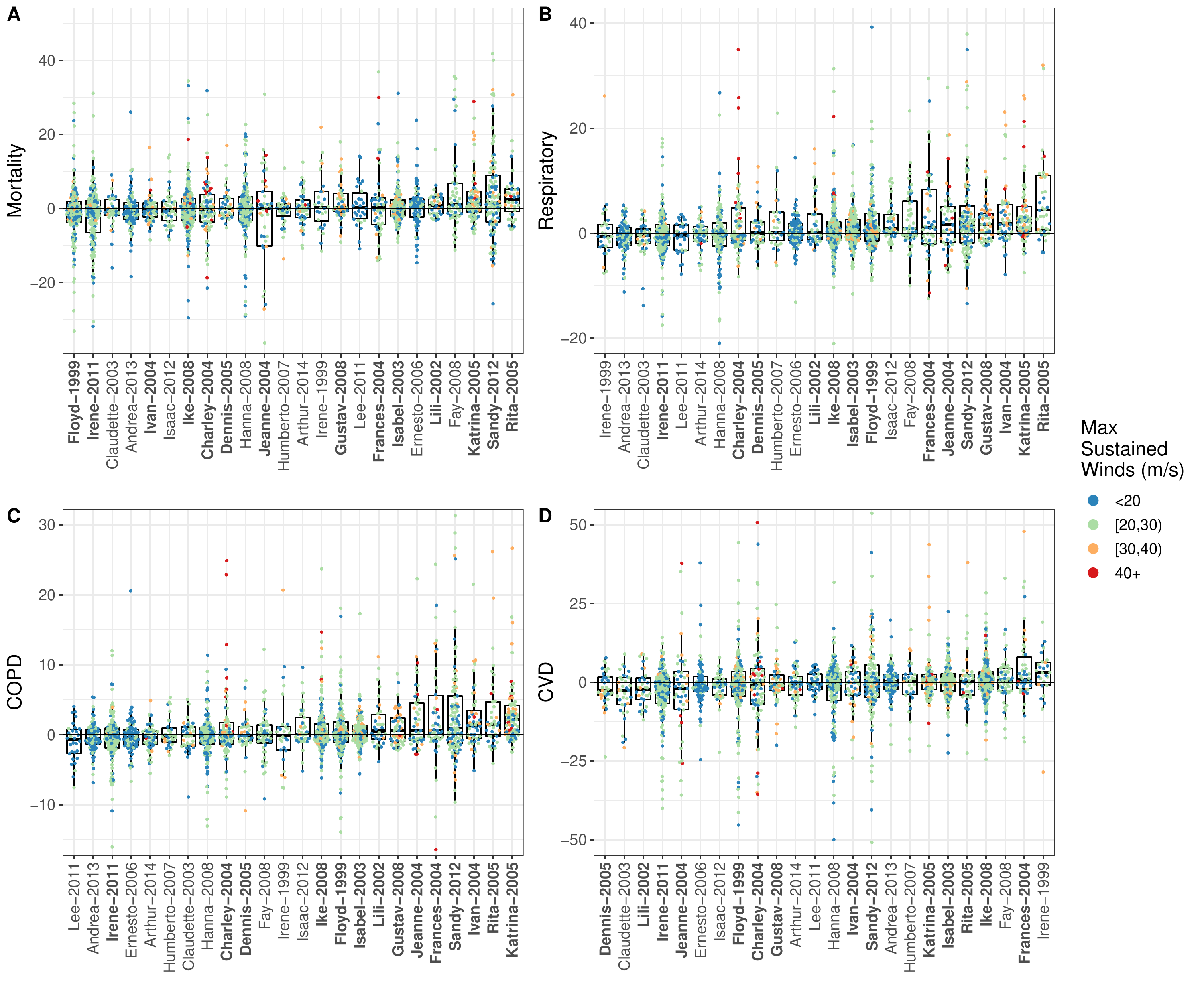}}
\caption{County-level individual excess events (IEE) estimates for mortality (A), respiratory hospitalizations (B), COPD hospitalizations (C), and CVD hospitalizations (D) for TCs that impacted $>25$ counties. The IEE is the estimated number of excess events in the county due to the TC. Distant outliers are cropped out for readability. Bolded TC labels indicate storm names that were subsequently retired-- retirement occurs when a TC is so destructive that re-using the name is considered to be insensitive \protect\citepSM{nhc2020}.}
\label{s:fig:excess_beeswarm}
\end{figure}

% latex table generated in R 3.6.3 by xtable 1.8-4 package
% Tue Sep  8 17:21:30 2020
\begin{table}[ht]
\centering
\caption{Predictive model coefficient posterior means (95\% CIs) from models with a natural cubic spline on sustained windspeed and year, not adjusted for state. Variable names with suffix \_s are components of a restricted cubic spline basis for the variable.}
\label{s:tab:pred_model_spl}
\footnotesize
\centerline{
\begin{tabular}{lcccc}
  \hline
 & Mortality & Resp & COPD & CVD \\ 
  \hline
(Intercept) & -115.89 (-378.81, 107.09) & 39.13 (-113.1, 198.02) & 18.07 (-117.22, 137.78) & -45.97 (-333.02, 243.32) \\ 
  vmax\_sust\_s1 & 4.34 (-7.09, 17.48) & -0.97 (-8.93, 6.6) & 0.43 (-5.6, 6.75) & 2.38 (-11.82, 15.2) \\ 
  vmax\_sust\_s2 & -92.36 (-248.23, 37.85) & -6.37 (-97.13, 87.88) & 7.56 (-69.23, 80.78) & -41.31 (-194, 129.42) \\ 
  vmax\_sust\_s3 & 181.19 (-46.76, 449.3) & 19.6 (-143.67, 180.82) & -12.29 (-140.88, 124.59) & 78.09 (-221.95, 345.87) \\ 
  poverty & 35.26 (-149.01, 206.88) & 114.78 (-22.65, 244.98) & 20.54 (-79.69, 119.58) & -121.81 (-375.06, 110.56) \\ 
  white\_pct & 14.23 (-38.78, 64.42) & -24.13 (-62.07, 12.89) & -32.12 (-61.3, -4.57) & 17.56 (-51.28, 89.96) \\ 
  owner\_occupied & -16.78 (-115.76, 84.09) & 17.94 (-60.87, 89.38) & 38.48 (-20.04, 95.68) & -91.02 (-239.35, 49.81) \\ 
  age\_pct\_65\_plus & -165.94 (-486.81, 139.45) & 7.66 (-241.24, 225.26) & 36.44 (-141.45, 213.65) & -95.8 (-514.39, 300.21) \\ 
  median\_age & 1.26 (-2.29, 5.05) & 0.02 (-2.27, 2.52) & -0.49 (-2.4, 1.57) & 1.72 (-2.57, 6.62) \\ 
  population\_density & -0.58 (-7.57, 6.61) & 0.02 (-5.34, 4.98) & 0.82 (-3.13, 4.8) & -0.97 (-9.88, 7.72) \\ 
  median\_house\_value & 2.83 (-7.67, 12.33) & 0.62 (-5.89, 7.82) & -0.38 (-6.05, 5.11) & -5.61 (-18.46, 7.29) \\ 
  no\_grad & 14.72 (-137.05, 154.17) & -62.79 (-167.04, 40.83) & -25.48 (-106.23, 58.74) & 36.94 (-143.44, 207.96) \\ 
  year\_s1 & -1.78 (-17.02, 14.35) & 6.97 (-4.91, 19.73) & 6.64 (-2.85, 15.56) & -2.08 (-23.65, 17.19) \\ 
  year\_s2 & -2.92 (-21.43, 13.73) & -11.03 (-24.57, 1.7) & -8.55 (-18.28, 1.86) & 4.73 (-16.43, 27.76) \\ 
  exposure & 1.19 (-0.78, 3.25) & -1.61 (-3.02, -0.17) & -1.29 (-2.35, -0.24) & 0.17 (-2.12, 2.84) \\ 
  sust\_dur & 0.01 (-0.02, 0.05) & 0.03 (0.01, 0.06) & 0 (-0.02, 0.02) & -0.01 (-0.05, 0.04) \\ 
  cc1 & -2.37 (-16.09, 12.75) & -1.24 (-11.75, 9.38) & -2.17 (-9.82, 5.43) & 8.19 (-9.89, 27.44) \\ 
   \hline
\end{tabular}}
\end{table}

% latex table generated in R 3.6.3 by xtable 1.8-4 package
% Mon Sep 14 14:20:32 2020
\begin{table}[ht]
\centering
\caption{Predictive model coefficient posterior means (95\% CIs) from models linear models (spline term only on the year variable), not adjusted for state. Variable names with suffix \_s are components of a restricted cubic spline basis for the variable.}
\footnotesize
\label{s:tab:pred_model_lin}
\begin{tabular}{lllll}
  \hline
 & Morality & Resp & COPD & CVD \\ 
  \hline
(Intercept) & -101.38 (-216.44, 8.75) & -6.18 (-84.45, 74.71) & 0.89 (-57.79, 60.08) & -19.05 (-155.97, 118.74) \\ 
  vmax\_sust & 3.21 (1.59, 5.09) & 1.39 (0.16, 2.59) & 1.36 (0.42, 2.28) & 0.84 (-1.24, 3.08) \\ 
  sust\_dur & -0.01 (-0.05, 0.02) & 0.02 (0, 0.04) & 0 (-0.01, 0.02) & -0.01 (-0.05, 0.03) \\ 
  exposure & 1.36 (-0.64, 3.3) & -1.59 (-2.96, -0.27) & -1.28 (-2.37, -0.2) & 0.27 (-2.08, 2.76) \\ 
  poverty & 48.39 (-135.71, 228.03) & 118.2 (-15.32, 241.09) & 22.36 (-82.63, 125.13) & -116.76 (-366.75, 110.8) \\ 
  white\_pct & 12.55 (-39.29, 64.74) & -25.91 (-66.17, 10.07) & -32.12 (-61.68, -3.08) & 17.14 (-50.02, 88.44) \\ 
  owner\_occupied & -4.82 (-110.78, 98.17) & 21.51 (-50.72, 97.63) & 38.45 (-16.56, 96.45) & -90.14 (-233.67, 52.13) \\ 
  age\_pct\_65\_plus & -123.03 (-446.07, 210.72) & 27.06 (-207.2, 253.55) & 44.02 (-133.55, 217.4) & -91.54 (-519.72, 298.64) \\ 
  median\_age & 0.78 (-2.99, 4.59) & -0.19 (-2.63, 2.31) & -0.53 (-2.54, 1.49) & 1.68 (-2.75, 6.38) \\ 
  population\_density & -1.1 (-7.94, 5.71) & -0.36 (-5.12, 5.01) & 0.75 (-2.95, 4.46) & -1.22 (-10.09, 7.51) \\ 
  median\_house\_value & 3.64 (-6.72, 13.2) & 0.84 (-6.38, 7.94) & -0.19 (-5.42, 5.52) & -5.52 (-18.1, 7.11) \\ 
  no\_grad & 19.87 (-126.09, 162.17) & -57.28 (-163.61, 47.03) & -26.03 (-108.24, 52.87) & 37.35 (-136.49, 216.22) \\ 
  year\_s1 & -0.85 (-17.39, 15.05) & 7.34 (-3.76, 18.99) & 6.72 (-3.06, 15.65) & -1.08 (-21.94, 19.25) \\ 
  year\_s2 & -3.84 (-21.69, 15.86) & -11.46 (-24.17, 1.48) & -8.8 (-18.86, 1.73) & 3.62 (-20.64, 27.44) \\ 
  cc1 & -1.23 (-16.32, 13.4) & -1.05 (-11.84, 9.45) & -2.47 (-10.36, 5.16) & 8.5 (-9.28, 27) \\ 
   \hline
\end{tabular}
\end{table}

\begin{figure}[h!]
\centering
\centerline{\includegraphics[scale=.9]{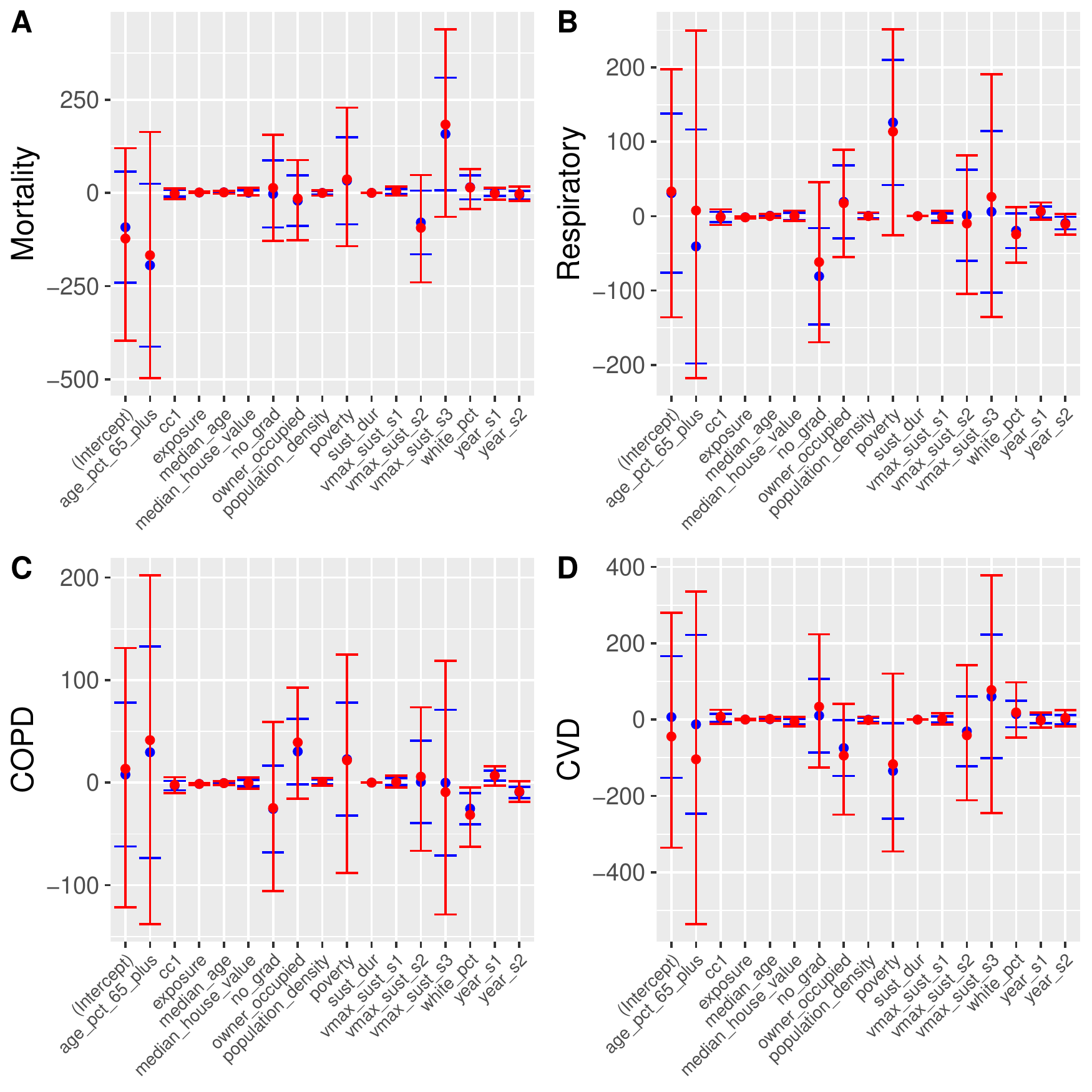}}
\caption{Point estimates and 95\% CIs from predictive models for each outcome that do propagate uncertainty from the causal models (red) and do not propagate uncertainty (blue).}
\label{s:fig:propagate}
\end{figure}

\clearpage

\bibliographystyleSM{chicago}

\bibliographySM{hurricanes_sm}

\end{document}